\documentclass[twocolumn]{aastex631}
\usepackage[utf8]{inputenc}
\usepackage{multirow}
\usepackage{float}
\usepackage{natbib}
\usepackage[]{threeparttable}
\usepackage{lineno}
\newcommand{\hi}{H\,{\sc i}}

\newcommand\kms{km$\,$s$^{-1}$}
\usepackage[percent]{overpic}
\usepackage{xcolor}            
\usepackage{placeins}

\shortauthors{Mutlu-Pakdil and Jones et al.}

\begin{document}

\title{SEAMLESS Survey: A Closer Look at Seven Local Volume Dwarf Galaxies in Diverse Environments}

\author[0000-0001-9649-4815]{Bur\c{c}in Mutlu-Pakdil}
\affil{Department of Physics and Astronomy, Dartmouth College, Hanover, NH 03755, USA}

\author[0000-0002-5434-4904]{Michael G. Jones}
\affil{IPAC, Mail Code 100-22, Caltech, 1200 E. California Blvd., Pasadena, CA 91125, USA}

\author[0000-0003-4102-380X]{David J. Sand}
\affil{Steward Observatory, University of Arizona, 933 North Cherry Avenue, Tucson, AZ 85721, USA}

\author[0000-0002-1763-4128]{Denija Crnojevi\'{c}}
\affil{Department of Physics \& Astronomy, University of Tampa, 401 West Kennedy Boulevard, Tampa, FL 33606, USA}

\author[0000-0001-5368-3632]{Laura C. Hunter}
\affiliation{Department of Physics and Astronomy, Dartmouth College, 6127 Wilder Laboratory, Hanover, NH 03755, USA}

\author[0000-0002-0956-7949]{Kristine Spekkens}
\affil{Department of Physics, Engineering Physics and Astronomy, Queen's University, Kingston, ON K7L 3N6, Canada}

\author[0000-0002-5177-727X]{Dennis Zaritsky}
\affil{Steward Observatory, University of Arizona, 933 North Cherry Avenue, Tucson, AZ 85721, USA}

\author[0000-0001-8354-7279]{Paul Bennet}
\affiliation{Space Telescope Science Institute, 3700 San Martin Drive, Baltimore, MD 21218, USA}

\author[0000-0002-3936-9628]{Jeffrey L. Carlin}
\affiliation{NSF–DOE Vera C. Rubin Observatory/NSF NOIRLab, 950 N. Cherry Ave., Tucson, AZ 85719, USA}

\author[0000-0001-7618-8212]{Richard Donnerstein}
\affil{Steward Observatory, University of Arizona, 933 North Cherry Avenue, Tucson, AZ 85721, USA}

\author[0000-0003-4379-6777]{Debasish Hazarika}
\affil{Steward Observatory, University of Arizona, 933 North Cherry Avenue, Tucson, AZ 85721, USA}

\author[0000-0002-7013-4392]{Donghyeon J. Khim}
\affil{Steward Observatory, University of Arizona, 933 North Cherry Avenue, Tucson, AZ 85721, USA}

\author[0000-0002-8217-5626]{Deepthi S. Prabhu}
\affil{Steward Observatory, University of Arizona, 933 North Cherry Avenue, Tucson, AZ 85721, USA}

\author{Cora Gillette}
\affil{Department of Physics and Astronomy, Dartmouth College, Hanover, NH 03755, USA}

\author{Grace Kelly}
\affil{Department of Physics \& Astronomy, University of Tampa, 401 West Kennedy Boulevard, Tampa, FL 33606, USA}

\author{Harshitha Madabhushani}
\affil{Department of Physics, University of Southern Florida, 4202 East Fowler Ave, ISA 2019, Tampa, FL 33620, USA}

\author{Annika Roberson}
\affil{Department of Physics and Astronomy, Dartmouth College, Hanover, NH 03755, USA}

\begin{abstract}
We present Hubble Space Telescope (HST) imaging of six Local Volume dwarf galaxies identified by the SEAMLESS survey -- Cetus~B, Hydrus~A, Ursa Major~C, Hydra~B, Grapes, and M83-dw14 -- along with Centaurus~C (Cen~C), discovered serendipitously in the M83-dw14 field. We report the discovery of M83-dw14 and Cen~C for the first time. Our HST imaging resolves each system into individual stars, enabling measurements of tip of the red giant branch (TRGB) distances (3.6-6.5~Mpc), structural parameters, and star formation histories (SFHs). We complement these data with new \hi\ observations from the Very Large Array for Ursa Major~C, Hydra~B, and M83-dw14 (also covering Cen~C), and from the Green Bank Telescope for Grapes, together with existing \hi\ limits for Cetus~B and Hydrus~A. With accurate TRGB distances, we place each target in its environmental context, ranging from a confirmed satellite (M83-dw14) to isolated systems (Hydrus~A, Hydra~B, Grapes, Cen~C). Cen~C ($M_V=-7.96$) is among the faintest isolated dwarfs known. It matches the size and luminosity of Leo~T, the lowest mass gas-rich galaxy currently known, yet unlike Leo~T it shows neither detected \hi\ ($\log(M_{\rm HI}/M_{\odot})<5.3$) nor recent star formation. Ursa Major~C shows strongly asymmetric, spatially offset \hi, suggesting that feedback may efficiently displace gas in such low-mass systems. Overall, our sample exhibits a broad range of SFHs and quenching states that does not track simply with present-day environment. This diversity is consistent with recent evidence that similarly varied SFHs occur even among isolated low-mass dwarfs, pointing to the important roles of inhomogeneous reionization, stochastic mass assembly, and internal feedback alongside environmental effects. 

\end{abstract}

\section{Introduction} \label{sec:intro}

Dwarf galaxies are unique laboratories for testing the predictions of cosmological models on the smallest scales, including the nature of dark matter and the physics of galaxy formation \citep{Bullock17,Sales2022}. However, two major challenges limit their use: (1) very few of the faintest galaxies are known beyond the Local Group, (2) the Local Group represents only one particular environment, making it difficult to draw robust cosmological and astrophysical conclusions from its satellite population alone. A systematic census of faint galaxies across a broad range of environments is therefore essential for understanding how dark matter and environment shape the formation and evolution of the lowest-mass galaxies. 

\begin{figure*}
\centering
\begin{overpic}[width=0.24\linewidth]{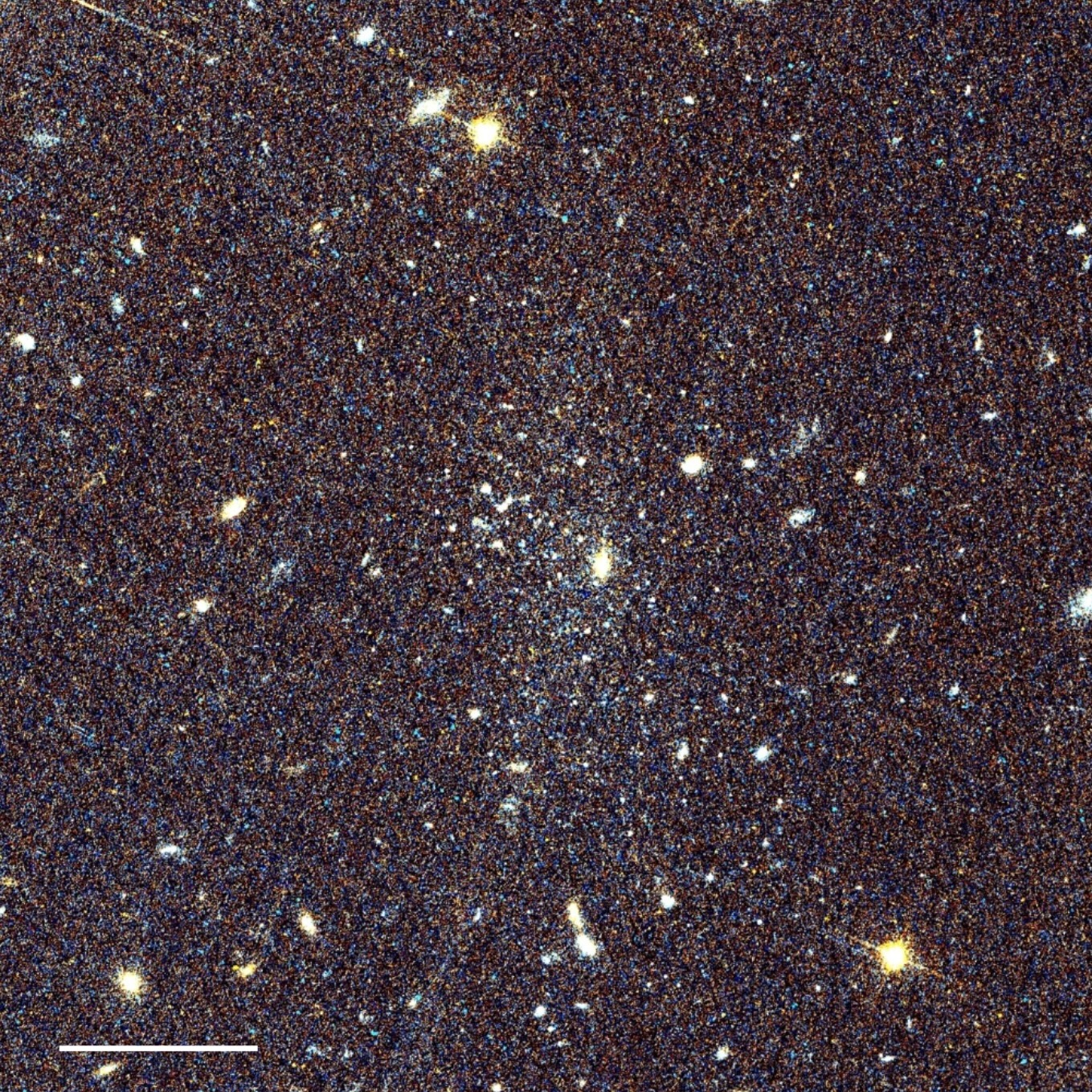}
  \put(5,92){\textbf{\textcolor{white}{Cetus B}}}
  \put(3,7){\textbf{\scriptsize\textcolor{white}{10 arcsec}}}
\end{overpic}
\begin{overpic}[width=0.24\linewidth]{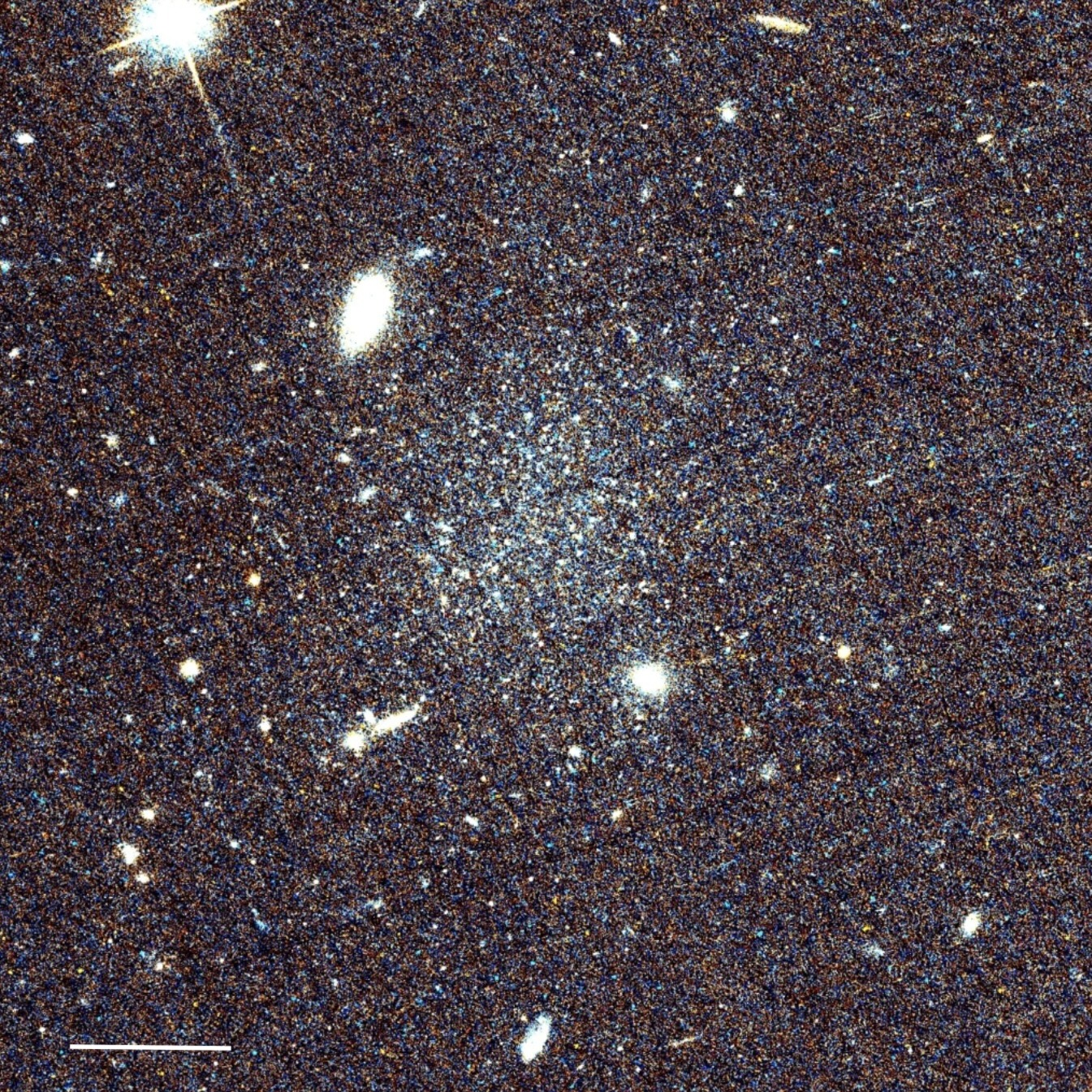}
  \put(5,92){\textbf{\textcolor{white}{Hydrus A}}}
  \put(3,7){\textbf{\scriptsize\textcolor{white}{10 arcsec}}}
\end{overpic}
\begin{overpic}[width=0.24\linewidth]{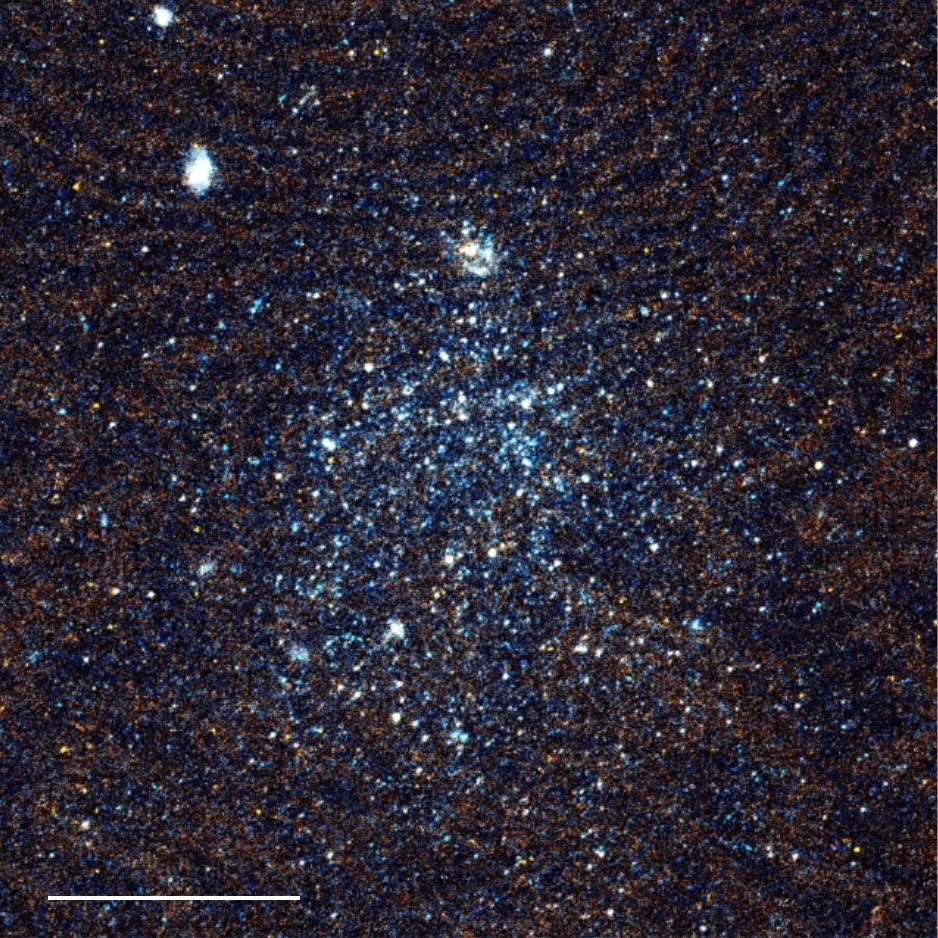}
  \put(5,92){\textbf{\textcolor{white}{Ursa Major C}}}
  \put(6,7){\textbf{\scriptsize\textcolor{white}{10 arcsec}}}
\end{overpic}
\begin{overpic}[width=0.24\linewidth]{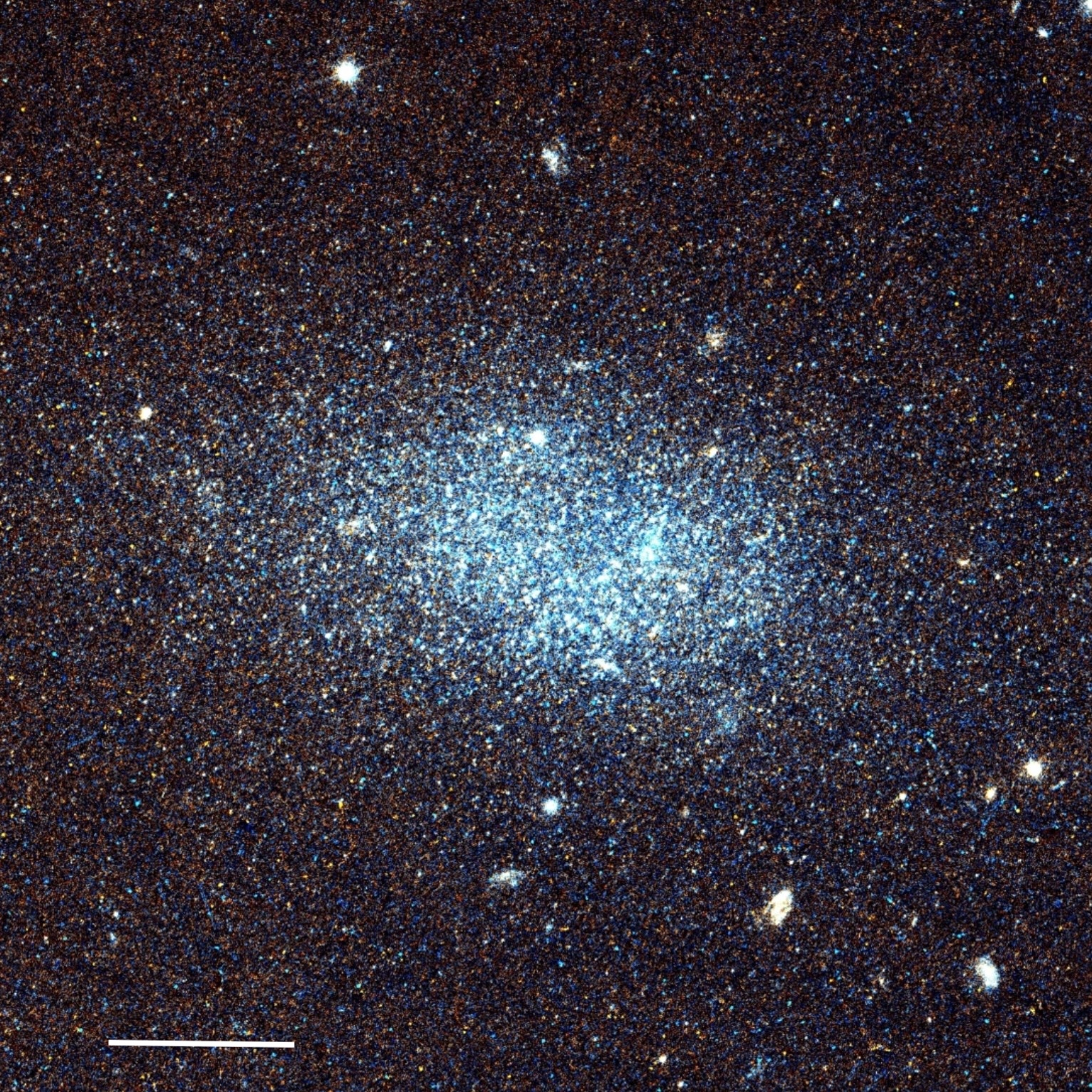}
  \put(5,92){\textbf{\textcolor{white}{Hydra B}}}
  \put(7,7){\textbf{\scriptsize\textcolor{white}{10 arcsec}}}
\end{overpic}\\[2mm]
\begin{overpic}[width=0.24\linewidth]{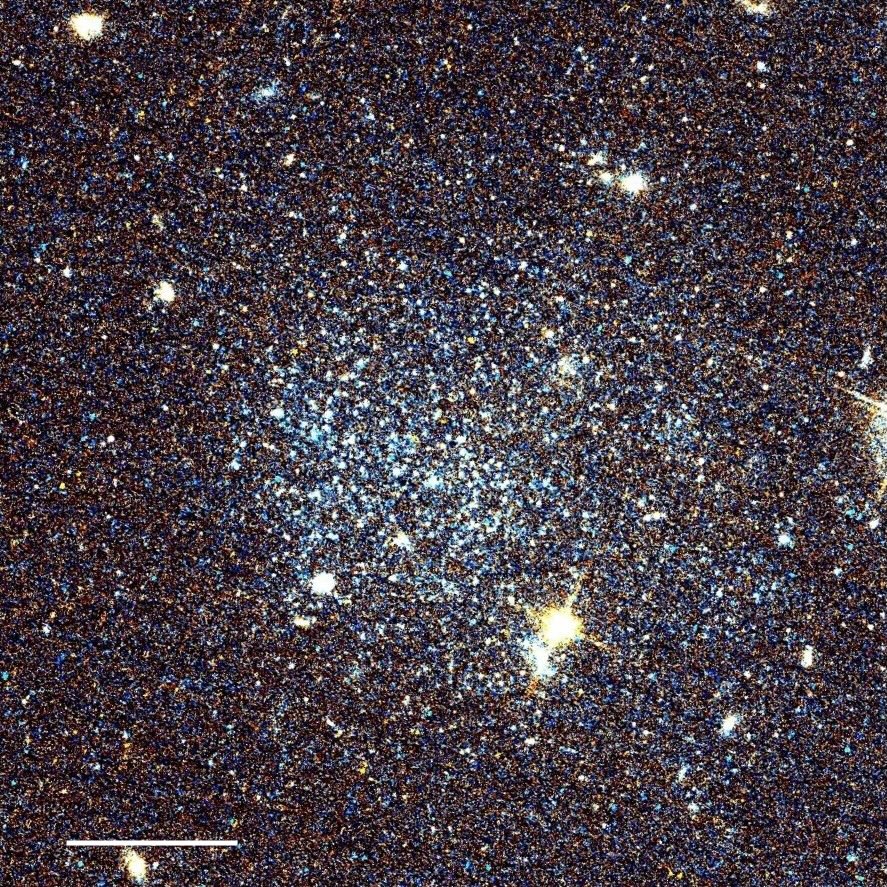}
  \put(5,92){\textbf{\textcolor{white}{Grapes}}}
  \put(7,7){\textbf{\scriptsize\textcolor{white}{10 arcsec}}}
\end{overpic}
\begin{overpic}[width=0.24\linewidth]{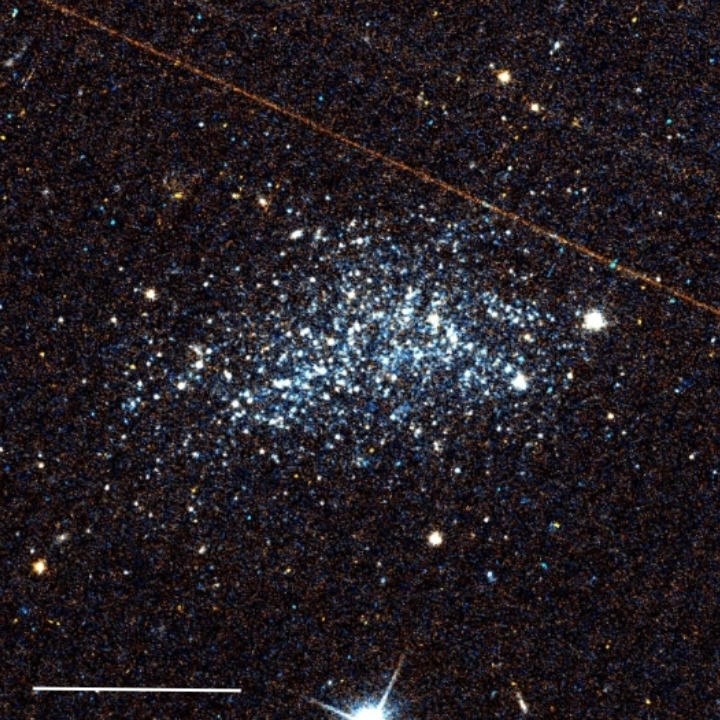}
  \put(5,92){\textbf{\textcolor{white}{M83-dw14}}}
  \put(7,7){\textbf{\scriptsize\textcolor{white}{10 arcsec}}}
\end{overpic}
\begin{overpic}[width=0.24\linewidth]{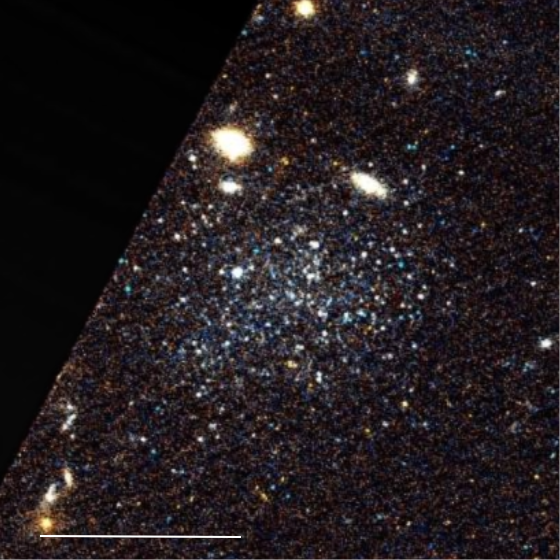}
  \put(5,92){\textbf{\textcolor{white}{Cen C}}}
  \put(13,7){\textbf{\scriptsize\textcolor{white}{10 arcsec}}}
\end{overpic}
\caption{Three-color HST images of our dwarfs. The images are created by combining the F606W band image (Blue), the average of F606W and F814W band images (Green), and the F814W band image (Red). North is up, and east is to the left. We note that Cen~C is a serendipitous discovery located within the field of M83-dw14.
\label{fig: image}}
\end{figure*}

The SEAMLESS (SEmi-Automated Machine LEarning Search for Semi-resolved galaxies) survey, first presented in \citet{Jones23}, is designed to uncover faint dwarf galaxies across all environments. SEAMLESS builds on the SMUDGes (Systematically Mapping Ultra Diffuse Galaxies; \citealt{Zaritsky2022}) pipeline, adapting its detection and machine-learning classification framework specifically to identify nearby ($\lesssim5$~Mpc) resolved and semi-resolved low-mass galaxies. Applied to Data Release 10 (DR10) of the DESI Legacy Imaging Surveys \citep{Dey2019}, it has already revealed numerous candidates, including the confirmed discoveries of Pavo \citep{Jones23}, Corvus~A \citep{Jones24}, Kamino \citep{BMP2026}, and Hydrus~A \citep{Fielder2025}. In this paper, we focus on six SEAMLESS detections (Cetus~B, Hydrus~A, Ursa Major~C, Hydra~B, Grapes, and M83-dw14)\footnote{We adopt constellation-based names for our targets, except for Grapes, for which we retain the name from the literature, and M83-dw14, which we name to reflect its association with M83 (see Section~\ref{sec:env}) and the 13 previously known satellites of this galaxy \citep{Muller2024}.}, together with Centaurus~C (Cen~C), which we discovered serendipitously in our follow-up HST data of M83-dw14. Cen~C was missed by our systematic SEAMLESS search because an imaging artifact lies directly on top of it in the DR10 imaging. 
Of these seven galaxies, M83-dw14 and Cen~C are reported here for the first time, while the remaining galaxies have been previously reported either by SEAMLESS or in other surveys and catalogs. We summarize their alternative designations and corresponding references in Table~\ref{tab:alternative_names} in Appendix~\ref{sec:names}.
Although several galaxies in our sample were previously identified, they lacked detailed characterization and accurate distance measurements, and very little was known about their physical properties or evolutionary histories. Here, we present HST imaging (see Figure~\ref{fig: image}) and neutral gas observations of all seven systems. The imaging is new for every galaxy except Grapes, for which we use archival data, enabling us to resolve each dwarf into individual stars. These observations allow us to measure robust tip of the red giant branch (TRGB) distances, structural parameters, and resolved star-formation histories, and to establish each system's three-dimensional environment. Interferometric imaging (4/7) and single-dish observations (3/7) detect neutral gas in four of these systems, providing their radial velocities for the first time along with the mass of their (generally) dominant baryonic component. Together, these seven systems span a wide range of environments, from confirmed satellites to genuinely isolated dwarfs, offering a valuable testbed for disentangling the environmental and internal processes that regulate star formation and quenching at the lowest galaxy masses.

In Section~\ref{sec:observations}, we describe the details of our HST data and photometry, as well as the \hi\ data used in this study.  In Section~\ref{sec:prop}, we measure the properties of our dwarfs, including their stellar populations, distance, structural parameters, gas content, and star formation histories. Finally, we compare our results with those of other Local Volume dwarf galaxies and Local Group satellites, and summarize our findings in Section~\ref{sec:discussion}.

\setcounter{table}{0}
\begin{table*}[htb]
\renewcommand{\thetable}{\arabic{table}}
\centering
\caption{Summary of HST Observations and Completeness.} \label{tab:obs}
\begin{tabular}{lccccc}
\tablewidth{0pt}
\hline
\hline
Name & UT Date &  Instrument/Filter & Exp  & 50\%  & 90\% \\
{}         & {}    & {}      & (s)  & (mag) & (mag) \\
\hline
Cetus~B   &  2025-06-21 & ACS/$F606W$ & 948  &  27.43 & 26.56    \\
{}     &  2025-06-21    & ACS/$F814W$ & 948  &  26.80 & 26.17    \\
Hydrus~A &  2025-09-13  & ACS/$F606W$ & 1026  & 27.70 & 26.56    \\
{}     &  2025-09-13    & ACS/$F814W$  & 1026  & 27.01 & 26.37    \\
Ursa Major C  &  2025-01-27 & WFC3/$F606W$ & 1150  &  27.25 & 26.19    \\
{}     &  2025-01-27    & WFC3/$F814W$ & 1150  &  26.51 & 26.03    \\
Hydra B  &  2025-12-17 & WFC3/$F606W$ & 1114  & 26.89  & 25.78    \\
{}     &  2025-12-17    & WFC3/$F814W$ & 1114  & 26.21  & 25.76    \\
Grapes  &  2020-10-15    & ACS/$F606W$ & 637  & 27.36  & 26.38    \\
{}     &  2020-10-15    & ACS/$F814W$ & 760  & 26.65  & 25.98    \\
M83-dw14 &  2025-03-15  & ACS/$F606W$   & 916  & 27.42 & 26.45    \\
{}     &  2025-03-15    & ACS/$F814W$  & 916  & 26.80 & 26.02    \\
\hline
\end{tabular}
\end{table*}

\section{Observations}
\label{sec:observations}

\subsection{HST Observations and Photometry} 

We obtained HST follow-up observations of Cetus~B, Hydrus~A (GO-17798, PI: Mutlu-Pakdil), and M83-dw14 (GO-17607, PI: Jones) using the Wide Field Channel (WFC) of the Advanced Camera for Surveys (ACS), and of Ursa Major~C and Hydra~B (GO-17891, PI: Sand) using the UVIS channel of the Wide Field Camera 3 (WFC3). Each target was observed for a total of one orbit in the F606W and F814W filters (see Table~\ref{tab:obs}), with each dwarf galaxy centered on one of the two chips. We also used archival ACS observations of Grapes, obtained as part of SNAP program 15922 (PI: R. B. Tully).  All the HST data used in this paper can be found in MAST: \dataset[10.17909/rbx4-nf65]{http://dx.doi.org/10.17909/rbx4-nf65}.
The HST images of all targets are shown in Figure~\ref{fig: image}.

We perform PSF photometry on pipeline-produced FLC images with \texttt{DOLPHOT} \citep{Dolphin2000}. We follow the recommended preprocessing steps and use the suggested input parameters from the \texttt{DOLPHOT} User Guide. The initial photometry is culled with the following criteria: the sum of the crowding parameters in the two bands is $<1$, the squared sum of the sharpness parameters in the two bands is $<0.1$, the signal-to-noise ratio is $>4$, and the object type is $\leq2$ in each band. We correct for MW extinction using the \citet{Schlegel98} reddening maps with the coefficients from \citet{Schlafly11}. Extinction-corrected photometry is used throughout this work. 

We perform artificial star tests to assess the photometric uncertainties and completeness of our observations. A total of $\sim$1,000,000 stars were added one at a time using the artificial star utilities in \texttt{DOLPHOT}, uniformly distributed both in color-magnitude space (i.e., $20$~$\leq$~F606W~$\leq$~$30$ and $-0.5$~$\leq$~F606W$-$F814W~$\leq$~$1.5$) and spatially across the field of view. Photometry and quality cuts were carried out in the same way as those applied to the original photometry. The 50\% and 90\% completeness limits for each field are listed in Table~\ref{tab:obs}.

\subsection{\hi\ Observations}

M83-dw14 was observed with the Karl G. Jansky Very Large Array (VLA) C-configuration in VLA-24A-308 (PI: M.~Jones) during March 2024, with a total on-source integration time of $\sim$2.5~h. Hydra~B and Ursa Major~C were both observed in March 2025 for VLA-25A-287 (PI: D.~Sand) with the VLA in D-configuration. Both had on-source integration times of $\sim$2.3~h. We had obtained a prior tentative detection of M83-dw14 with the Robert C. Byrd Green Bank Telescope (GBT), and therefore used a narrow 4~MHz bandwidth (with 3.9~kHz channels) centered on the \hi \ line shifted to a radial velocity of $\sim$500~\kms. Hydra~B and Ursa Major~C had no prior \hi \ detections, and those observations therefore used a wider 16~MHz band with 5.2~kHz channels.

These VLA observations were reduced using the pipeline developed by \citet{Jones+2023a}, which uses standard data reduction procedures from the Common Astronomy Software Applications package \citep[CASA;][]{CASA}. Radio frequency interference (RFI) was flagged with a combination of manually tailored flags and automated flagging with CASA tasks \texttt{tfcrop} and \texttt{rflag}. Ursa Major~C and M83-dw14 were both minimally impacted by RFI, with each losing approximately 15\% of their on-source data. However, there was severe RFI present throughout the observations of Hydra~B, which resulted in about 55\% of the on-source data being flagged. The continuum was subtracted using a first-order fit to the line-free channels. The data were then smoothed to a spectral resolution of 5~\kms \ (to improve S/N) before imaging with a Briggs robust parameter of 2.0. A summary of the properties of these VLA observations is given in Table~\ref{tab:HIobs}.

Note that Cen~C was also contained within the primary beam of the observations of M83-dw14. Thus, the same data set was used to investigate its \hi \ properties.

Grapes was observed with the GBT as part of GBT26A-555 (PI: M.~Jones) in January 2026. It was observed in 600-second ON/OFF pairs (300~s ON, 300~s OFF) with a total on-source integration time of 3600~s.  We used the L-band receiver coupled with the Versatile GBT Astronomical Spectrometer (VEGAS) receiver operating in Mode 10, with a 23.44~MHz bandwidth and a 0.7~kHz channel width (0.15 \kms) to cover a velocity range from $-$500 to 4500~\kms. The data were reduced using the GBTIDL routine \texttt{getps}, followed by baseline subtraction and spectral smoothing to enhance the S/N ratio. The data were flux calibrated using the default L-band noise diode and multiplied by 1.2 to account for the calibration offset identified in \cite{Goddy2020}.

\begin{table}
\centering
\caption{Properties of VLA observations} 
\begin{tabular}{lccc}
\hline \hline
 & Ursa Major C & Hydra B  & M83-dw14 \\ \hline
$\sigma_\mathrm{rms}$/mJy & 0.8 & 1.4 & 1.0  \\
$\Delta v$/\kms & 5.0 & 5.0 & 5.0 \\
Beam size & 58\arcsec $\times$ 51\arcsec & 104\arcsec $\times$ 55\arcsec & 44\arcsec $\times$ 15\arcsec \\
$N_\mathrm{HI}/\mathrm{cm}^{-2}$ & $6.3 \times 10^{18}$ & $5.9 \times 10^{18}$ & $3.6 \times 10^{19}$ \\ \hline
\end{tabular}
 \begin{tablenotes}
      \small
      \item $\sigma_\mathrm{rms}$: rms noise of the cleaned data cube. $\Delta v$: Channel resolution after imaging. Beam size: Synthesized beam size. $N_\mathrm{HI}$: 3$\sigma$ \hi \ column density sensitivity over 10~\kms.
 \end{tablenotes}
 \label{tab:HIobs}
\end{table}

\section{Properties of Our Targets \label{sec:prop}}

\subsection{Color-Magnitude Diagram \label{sec:cmd}}
\begin{figure*}
\centering
\includegraphics[width = \linewidth]{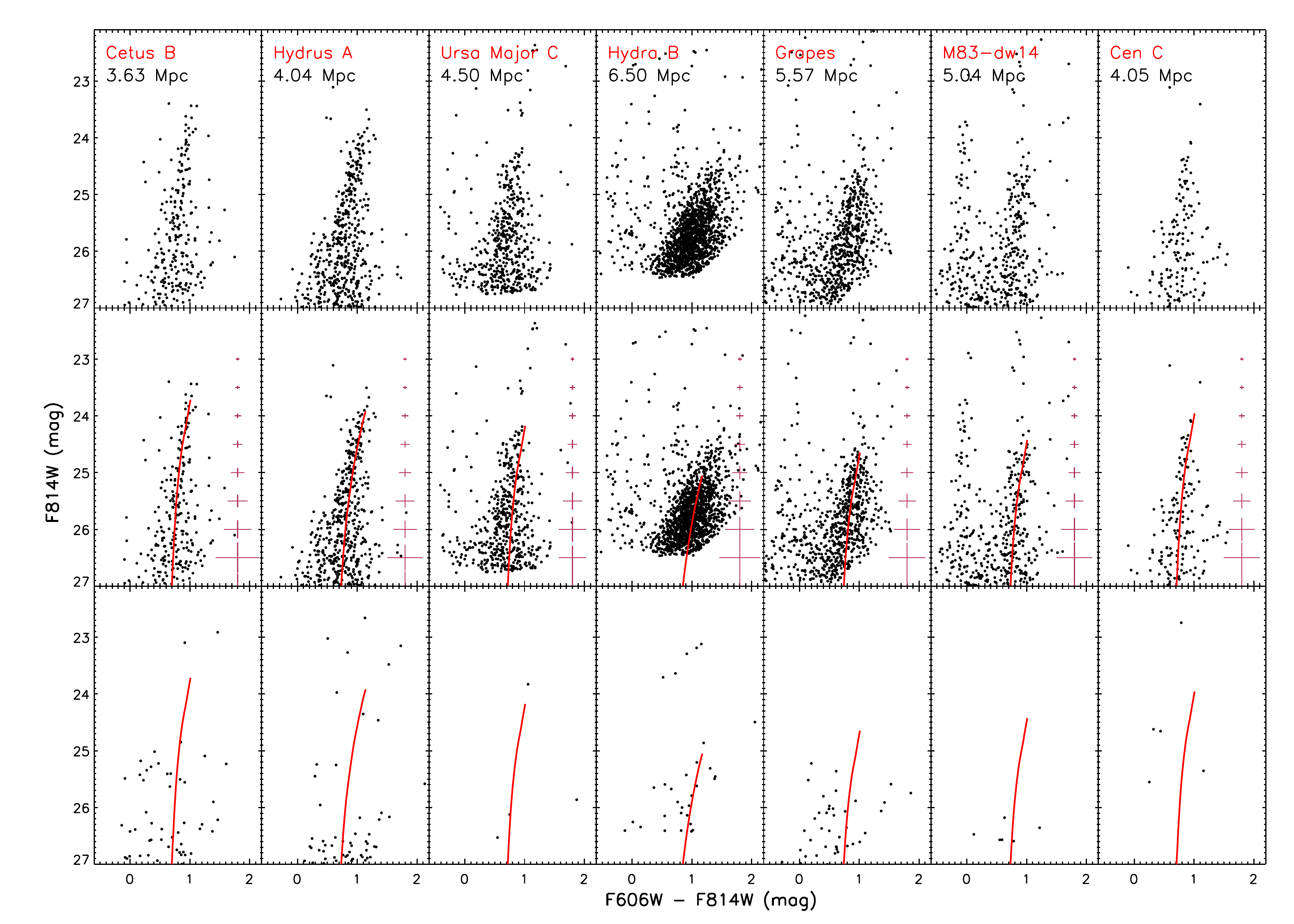}
\caption{Top: HST CMDs showing the stars within $\sim$2$\times$$r_h$ (see Table~\ref{tab:dwarfs}) of each dwarf galaxy: Cetus~B, Hydrus~A, Ursa Major~C, Grapes, M83-dw14, and Cen~C. Middle: same as the top panel, but with the relevant old (13 Gyr) isochrones overlaid for each dwarf galaxy. We adopt [M/H] = $-1.5$ for Hydrus~A and Hydra~B, and [M/H] = $-2.2$ for the remaining dwarfs. Maroon error bars show the mean photometric errors from artificial stars. Bottom: CMDs of representative field regions of equal area, selected from the chip that does not contain the target dwarf galaxy. \label{fig:cmd} }
\end{figure*}

Figure~\ref{fig:cmd} shows the HST CMDs, which include stars within two half-light radii (2$\times$$r_h$, see Table~\ref{tab:dwarfs} and Section~\ref{sec:str}). Overplotted in red on each CMD is a 13~Gyr PARSEC isochrone \citep{Bressan2012}, with [M/H] = $-1.5$ for Hydrus~A and Hydra~B, and [M/H] = $-2.2$ for the remaining dwarfs. Maroon error bars show the mean photometric errors from artificial stars, and are plotted at an arbitrary color for convenience. Each dwarf is clearly resolved into its constituent red giant branch (RGB) stars in the HST data, and shows an old, metal-poor stellar population. The bottom panels show CMDs of representative field regions of equal area, selected from the chip that does not contain the target dwarf galaxy. These CMDs serve as a typical field area in the vicinity of the dwarf, helping to assess field contamination. 

\begin{figure*}
\centering
\includegraphics[width = 0.9\linewidth]{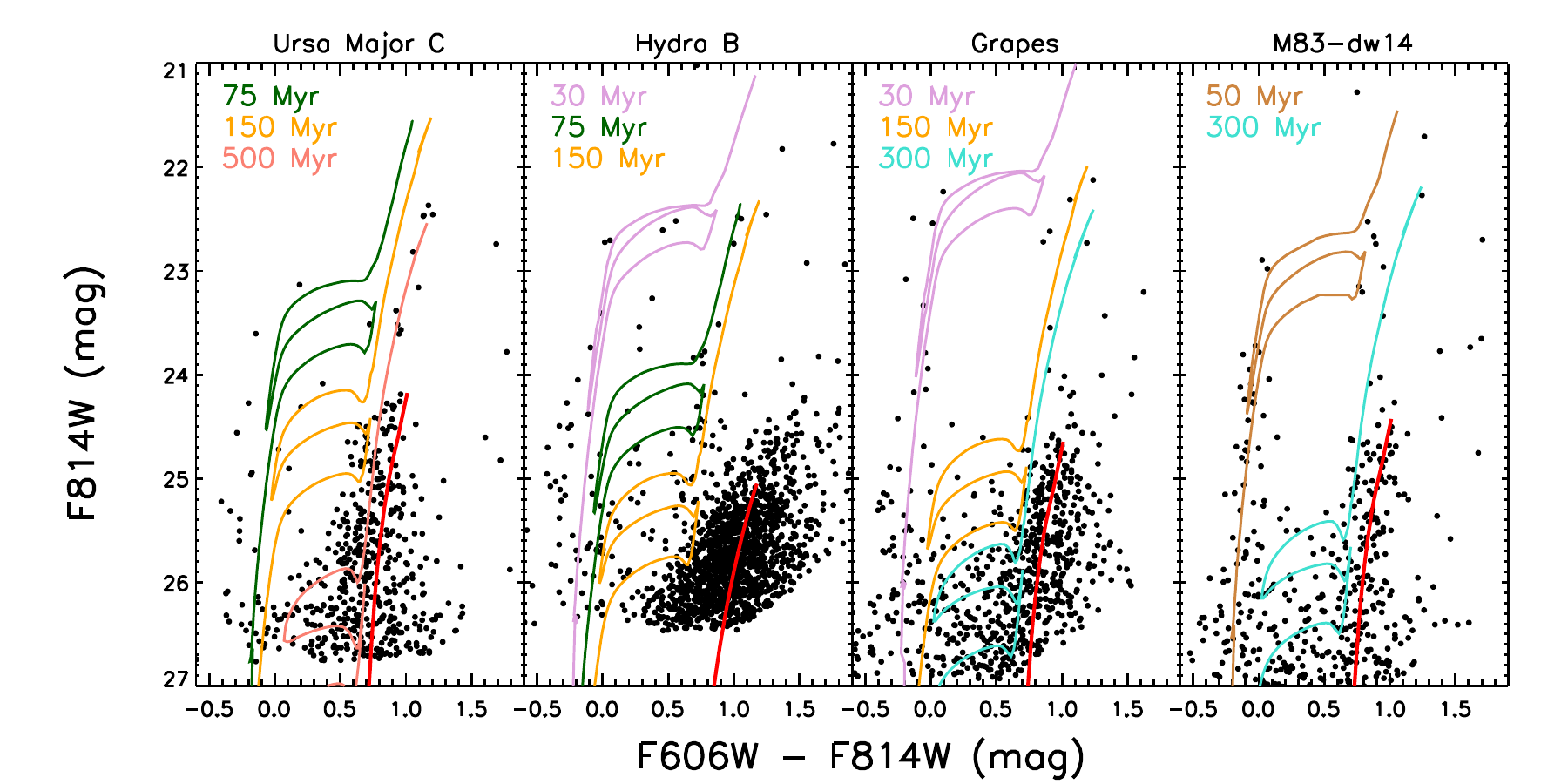}
\caption{HST CMDs of Ursa Major~C (left), Hydra~B (middle-left), Grapes (middle-right), and M83-dw14 (right), shown as in Figure~\ref{fig:cmd} but with an emphasis on the blue stellar populations. PARSEC isochrones are overlaid to highlight young stellar populations. Younger isochrones assume [M/H] = $-1.5$, with ages indicated in the legend. \label{fig:cmd2}}
\end{figure*}

Ursa Major~C, Hydra~B, Grapes and M83-dw14 also contain several blue stars with F606W$-$F814W~$<$~$0.5$. Figure~\ref{fig:cmd2} shows the same CMDs as Figure~\ref{fig:cmd}, but with PARSEC isochrones overlaid for a range of stellar population ages. The younger isochrones assume a metallicity of [M/H] = $-1.5$, with ages indicated in the legend. We identify young stellar populations by looking for a qualitative match between these isochrones and stars bluer than F606W$-$F814W~$=~0.5$. This comparison suggests the presence of young stellar populations with approximate ages of 75~Myr in Ursa Major~C, 30~Myr in Hydra~B and Grapes, and 50~Myr in M83-dw14. Detailed star formation history measurements for these systems are presented in Section~\ref{sec:sfh}.

\subsection{Distance \label{sec:dist}}

We measure distances to our targets using the TRGB method \citep[e.g.,][]{Lee1993,Salaris2002,Rizzi2007}, as described in \citet{Crnojevic19} and \citet{BMP2022}. We first apply a color correction to our photometry, to reduce the curvature and width of the RGB and increase the sharpness in the break of the luminosity function, following \cite{jang17} (their formula~5 and Table~6); we then compute the observed luminosity function for RGB stars, applying a color cut of F606W$-$F814W~$>~0.6$ to avoid any contamination from possible young populations (see Figure~\ref{fig:cmd}). The luminosity function is fit with a model that is convolved by the appropriate photometric uncertainty and completeness function as derived from our artificial star tests. Our final uncertainties combine the fitting uncertainties (which include the artificial star test results), the uncertainties from the TRGB zeropoint calibration and the applied color correction, and an assumed 10\% uncertainty on the adopted extinction value, added in quadrature. The TRGB values, the distance moduli, and the distances for our targets are reported in Table~\ref{tab:dwarfs}. 

For Cen~C, the paucity of RGB stars prevented the code from converging to a reasonable result, similar to the challenges encountered when analyzing comparably faint dwarf galaxies (e.g., \citealt{BMP2022}). We thus apply a simple Sobel filter edge-detection algorithm to this dwarf (following the prescription from \citealt{sakai96}), and compute the related uncertainties with a Monte Carlo (MC) calculation, varying the position of the stars in the CMD within their photometric errors.    

\subsection{Structural Properties \label{sec:str}}

We derive structural parameters of our targets using the maximum-likelihood (ML) method described by \citet{Sand12}, which builds on the approach of \citet{Martin08}. We use stars consistent with an old, metal-poor isochrone (13 Gyr, [M/H]~=~$-2.0$ for all systems except Hydrus~A and Hydra~B, for which we use [M/H]~=~$-1.5$) in color-magnitude space after taking into account photometric uncertainties, within our 90\% completeness limits. When photometric errors are less than 0.1~mag, we inflate the uncertainty to 0.1~mag to select stars for our ML analysis. We fit a standard exponential profile plus a constant background to the data. The resulting structural parameters are summarized in Table~\ref{tab:dwarfs}. We determine uncertainties by bootstrapping the data 1,500 times and recalculating the structural parameters for each sample.

\subsection{Luminosities \label{sec:lum}}
We derive absolute magnitudes by using the same procedure as in \citet{BMP2022}, as was first described in \citet{Martin08}. Artificial CMDs are generated using a 13~Gyr PARSEC isochrone with [M/H]~=~$-2.0$, except for Hydrus~A and Hydra~B, for which we adopt [M/H]~=~$-1.5$. We use the corresponding luminosity functions assuming a Kroupa IMF \citep{Kroupa2001}, and incorporate the measured completeness and photometric uncertainties. We randomly select the same number of stars from these CMDs to match the observed stellar counts, and sum their fluxes while extrapolating the contribution from unresolved stars using the luminosity function. The quoted magnitudes and uncertainties reflect 1000 realizations, with additional contributions from Poisson uncertainties in the stellar counts added in quadrature.

For Ursa Major~C, Hydra~B, M83-dw14, and Grapes, we also account for the luminosity contribution of their young and intermediate-age stellar populations. We select stars with F814W~$<$~$26$~mag and $-0.5$~$<$~F606W$-$F814W~$<$~$0.3$ within 2$\times$$r_{h}$, and calculate the total flux from these stars. The same selection is applied in dwarf-free regions of the field of view to estimate and subtract background contamination. Given the overall agreement between these blue stars and the youngest isochrones shown in Figure~\ref{fig:cmd2} (with ages of 75, 30, 30, and 50 Myr for Ursa Major~C, Hydra~B, Grapes, and M83-dw14, respectively)\footnote{We note that adopting other young isochrones of slightly different ages and their corresponding luminosity functions does not change our results within the uncertainties.}, we adopt the corresponding PARSEC luminosity function and extrapolate the flux of unaccounted young stars. We then consider stars brighter than the TRGB with $0.5$~$<$~F606W$-$F814W~$<$~$1.5$ within 2$\times$$r_{h}$, and compute the total flux from these stars. Background stars are estimated and subtracted using the same approach for this selection as well. The sum of the luminosities from old RGB stars and younger stellar populations is adopted as the total absolute magnitude for each dwarf (see Table~\ref{tab:dwarfs}). 

\begin{table*}
\centering
\small
\caption{Properties of our targets \label{tab:dwarfs}}
\begin{tabular}{lccccccc}
\tablewidth{0pt}
\hline
\hline
Parameter   & Cetus~B & Hydrus~A & Ursa Major C & Hydra B & Grapes  & M83-dw14 & Cen~C\\
\hline
R.A. (deg)  & $10.145485$  &  $56.194692$ & $203.35701$ & $197.44038$ & $178.02314$ & $204.74115$ & $204.73214$ \\
Dec. (deg)  & $-20.556725$ &  $-70.354717$ & $56.500077$ & $-23.542465$ & $54.792785$ & $-32.018012$ & $-31.989991$ \\
F814W$_{\rm TRGB}$ (mag) & $23.78\pm0.22$ & $24.02\pm0.12$ & $24.25\pm0.14$ & $25.05\pm0.05$ & $24.71\pm0.04$ & $24.50\pm0.06$ & $24.02\pm0.21$ \\
$m-M$ (mag) &  $27.80\pm0.23$    &   $28.03\pm0.14$ & $28.26\pm0.15$ & $29.06\pm0.08$ & $28.73\pm0.07$ & $28.51\pm0.08$  & $28.04\pm0.22$ \\  
$D$ (Mpc)  &   $3.63\pm0.40$    &    $4.04\pm0.26$  & $4.50\pm0.30$  & $6.50\pm0.23$ & $5.57\pm0.19$ & $5.04\pm0.20$ & $4.05\pm0.40$ \\ 
$M_{V}$   & $-8.24\pm0.13$ & $-9.31\pm0.09$ & $-9.64\pm0.09$ & $-11.96\pm0.11$ & $-10.79\pm0.07$ & $-9.98\pm0.16$ & $-7.96\pm0.17$   \\
$E(B-V)$ & 0.022 & 0.046 & 0.009 & 0.127 & 0.009 & 0.051 & 0.053\\
$\log(M_\ast/M_{\odot})$  & $5.44^{+0.09}_{-0.10}$ & $5.79^{+0.05}_{-0.05}$ & $5.66^{+0.09}_{-0.10}$ & $6.95^{+0.02}_{-0.03}$ & $6.17^{+0.06}_{-0.07}$ & $5.63^{+0.07}_{-0.08}$ & $5.23^{+0.10}_{-0.11}$ \\
$\log(M_\mathrm{HI}/M_{\odot})$ & $<5.26$$\dagger$  & $<6.42$$\dagger$ & $5.76\pm0.11$ & $6.72\pm0.08$ & $6.78\pm0.10$ & $6.38\pm0.07$ & $<5.30\ddagger$ \\
$v$ (\kms) & \nodata & \nodata & $150\pm2$ & $739\pm2$ & $226\pm1$ & $518\pm1$ & \nodata \\    
$W_{50}$ (\kms) & \nodata & \nodata & $25\pm3$ & $44\pm2$ & $20.7\pm1.5$ & $14\pm2$ & \nodata \\
$S_{\mathrm{HI}}$ (Jy~\kms) & $<0.058$$\dagger$ & $<0.69$$\dagger$ & $0.12\pm0.03$ & $0.53\pm0.09$ & $0.83\pm0.16$ & $0.40\pm0.07$ & $<0.052$ \\
$r_{h}$ (arcsec)     & $19\pm3$ & $16\pm1$ & $14\pm1$ & $28\pm1$ & $18\pm1$ & $10\pm1$ & $7\pm1$ \\ 
$r_{h}$ (pc)   	     & $285\pm45$ & $309\pm19$  & $290\pm21$ & $939\pm34$ & $486\pm27$ & $227\pm23$ & $137\pm20$ \\  
$\epsilon$           & $0.64\pm0.06$ & $0.31\pm0.07$ & $0.32\pm0.06$ & $0.54\pm0.02$ & $0.23\pm0.06$ & $0.51\pm0.06$ & $0.40\pm0.10$  \\ 
P.A. (deg) & $16\pm3$ & $-15\pm5$ & $17\pm7$ & $74\pm2$ & $22\pm10$ & $103\pm6$ & $115\pm10$ \\

\hline
\end{tabular}
  \begin{tablenotes}
      \small
      \item  R.A.: the Right Ascension (J2000.0). Dec.: the Declination (J2000.0). $m-M$: the distance modulus. $D$: the distance of the galaxy in Mpc. $M_{V}$: the absolute V-band magnitude. $E(B-V)$: Foreground reddening. $\log(M_{\star}/M_{\odot})$: the stellar mass in solar mass, derived from SFH (see Section~\ref{sec:sfh}). $\log(M_\mathrm{HI}/M_{\odot})$: the \hi~mass of each object or their $3\sigma$ upper limits. $v$: radial velocity. $W_{50}$: velocity width at 50\% level of the \hi \ line profile. $S_\mathrm{HI}$: \hi \ line integrated flux. $r_{h}$: the elliptical half-light radius along the semi-major axis. $\epsilon$: ellipticity which is defined as $\epsilon=1-b/a$, where $b$ is the semiminor axis and $a$ is the semimajor axis. P.A.: position angle.  
      \item ($\dagger$) \hi \ limits taken from \citet{Fielder2025} and adjusted for our updated distance measurements.
      \item ($\ddagger$) This \hi \ limit assumes that the unknown radial velocity of Cen~C falls in the approximate range 100-900~\kms (see discussion in Appendix~\ref{sec:CenC_spec}).
    \end{tablenotes}
\end{table*}

\subsection{Gas Content}

\begin{figure*}[!tb]
    \centering
    \includegraphics[width = 0.312\linewidth]{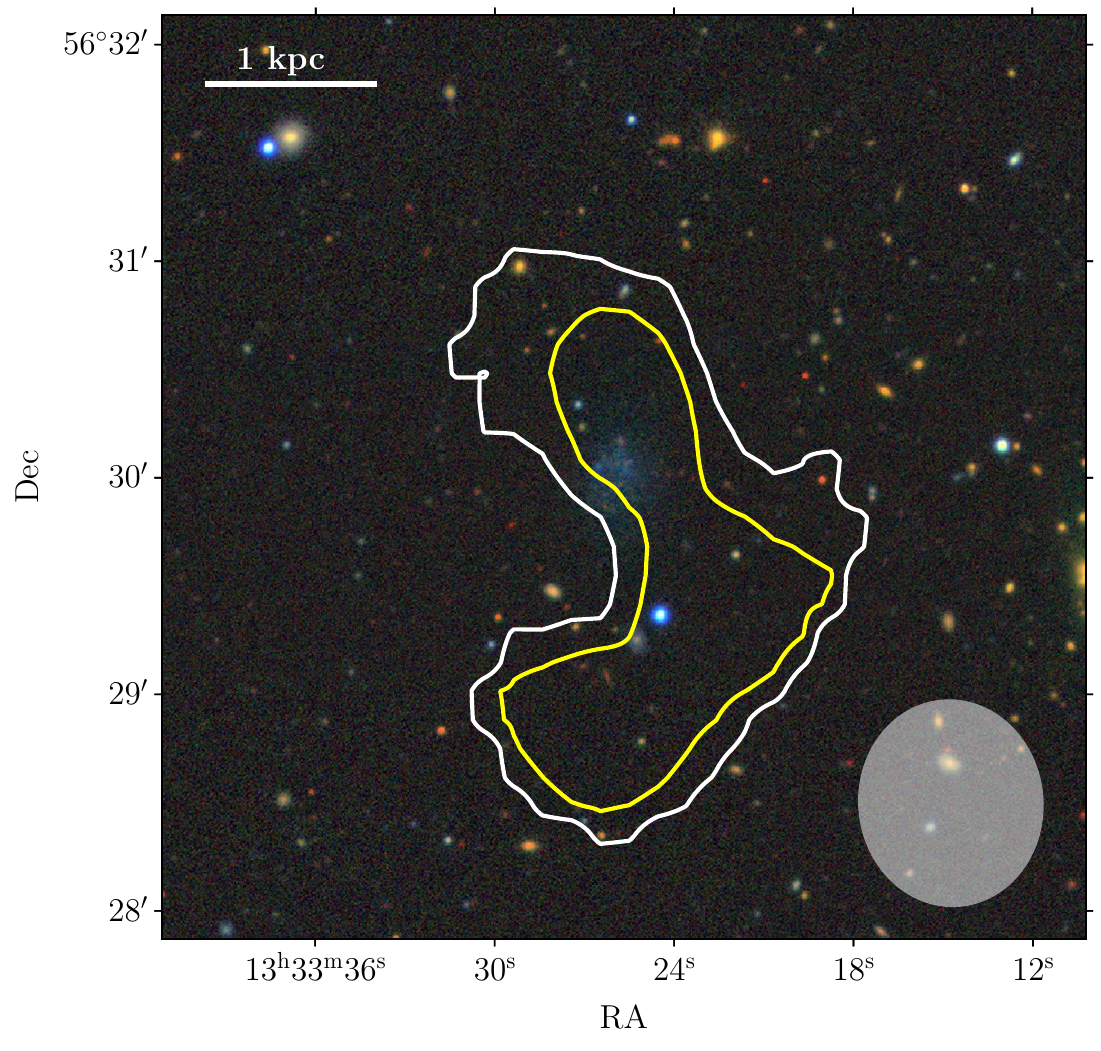}
    \includegraphics[width = 0.32\linewidth]{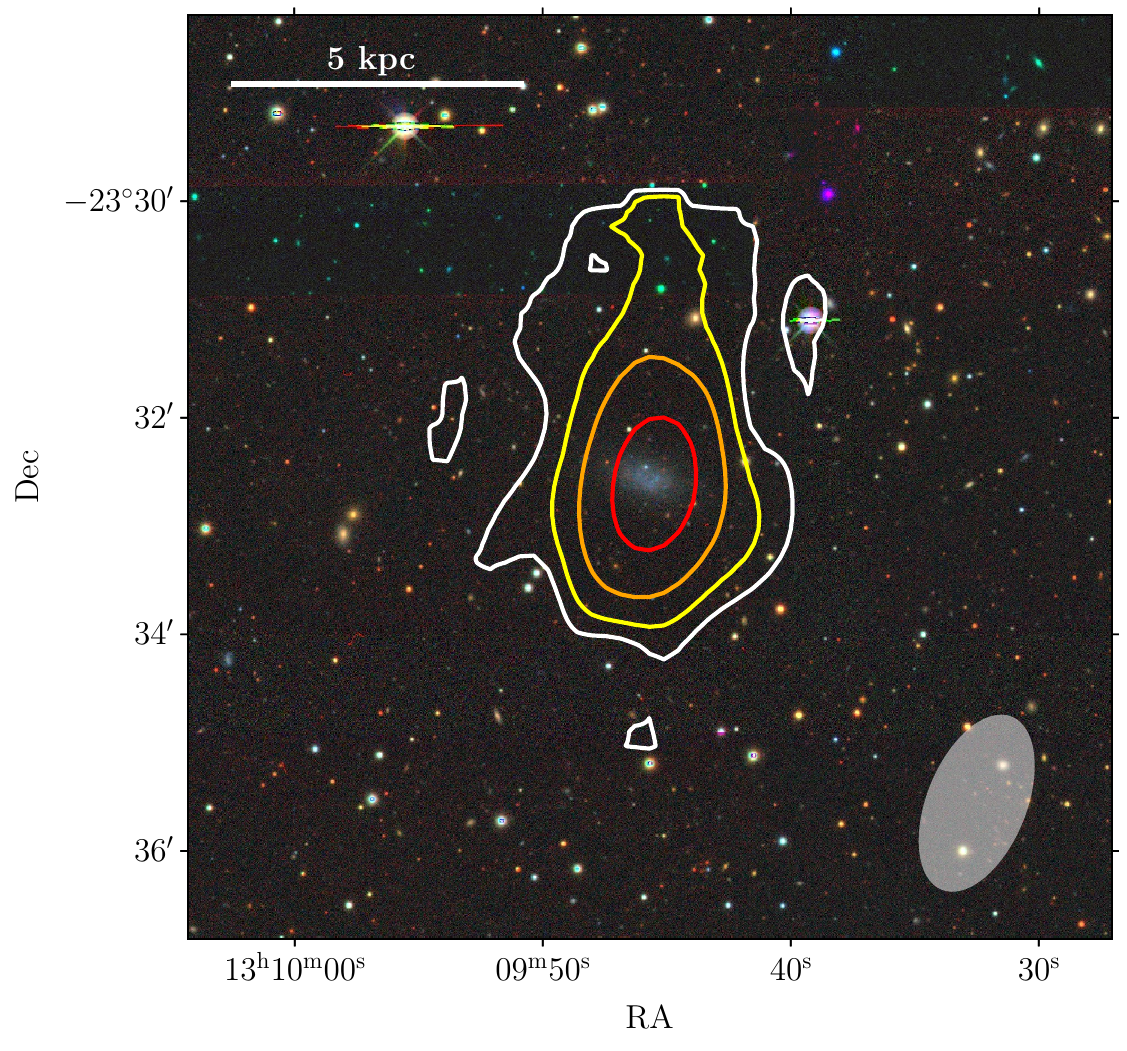}
    \includegraphics[width = 0.32\linewidth]{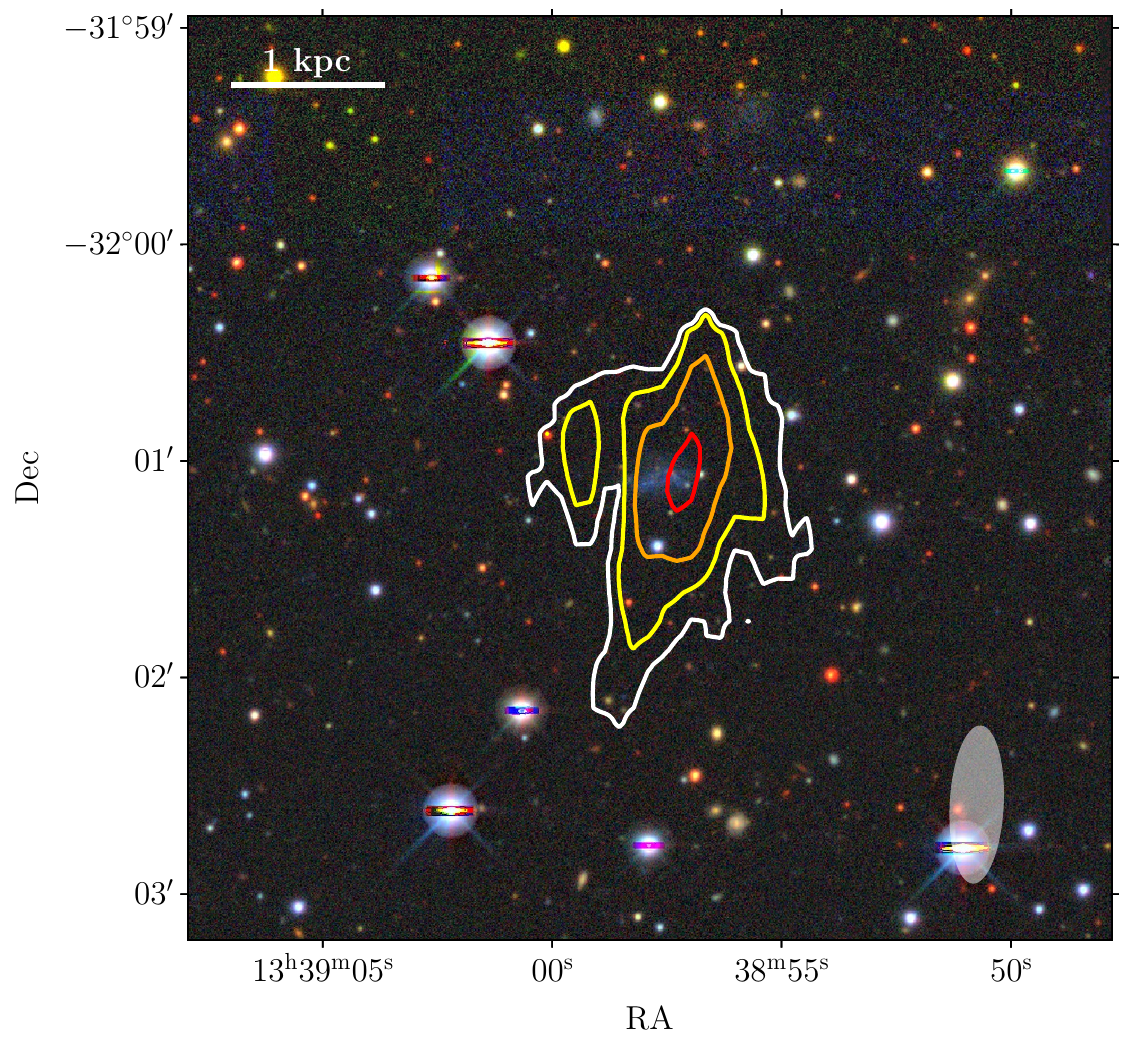} \\
    \includegraphics[width = 0.314\linewidth]{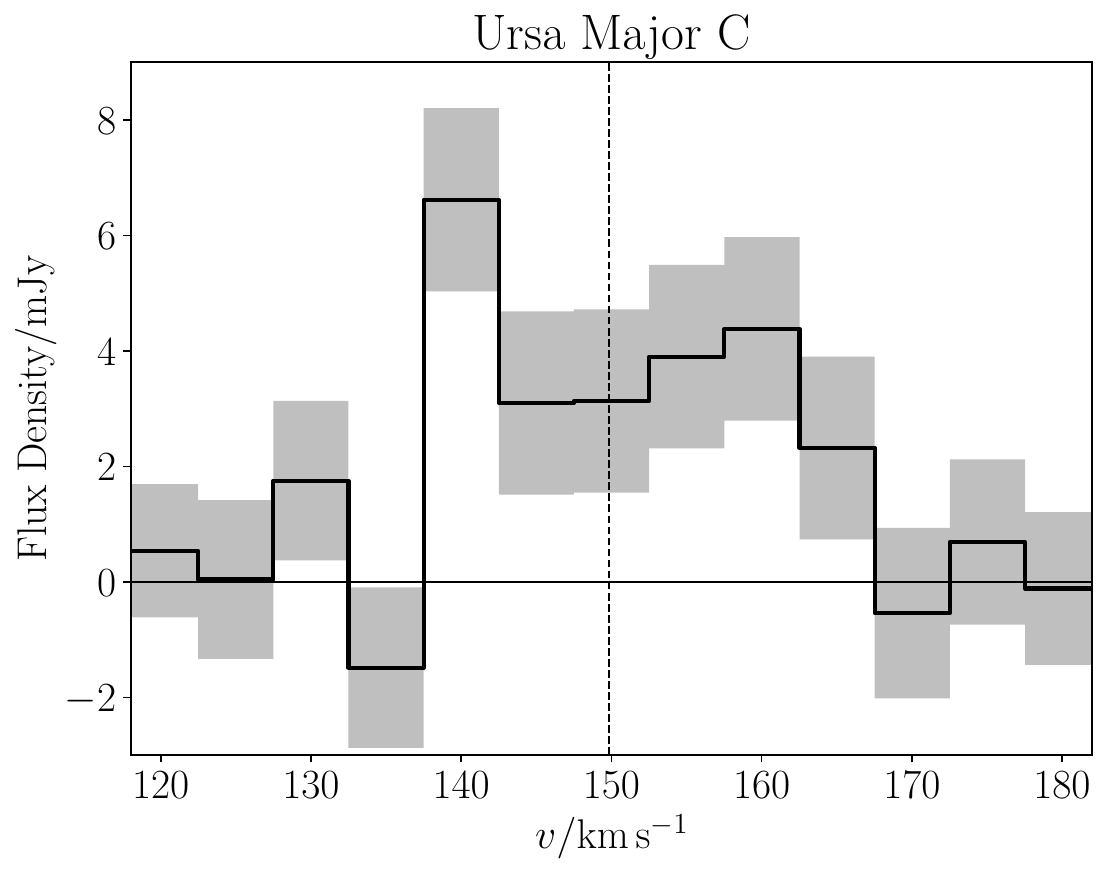}
    \includegraphics[width = 0.32\linewidth]{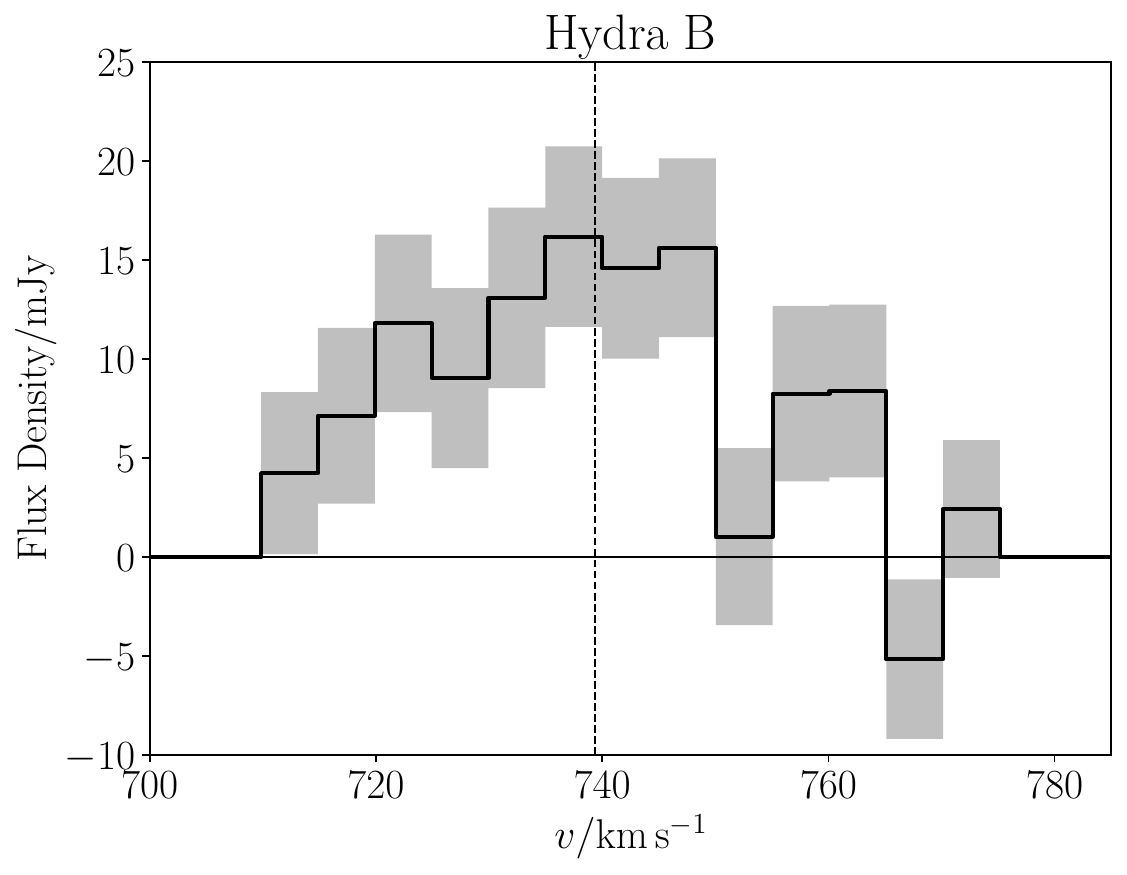}
    \includegraphics[width = 0.32\linewidth]{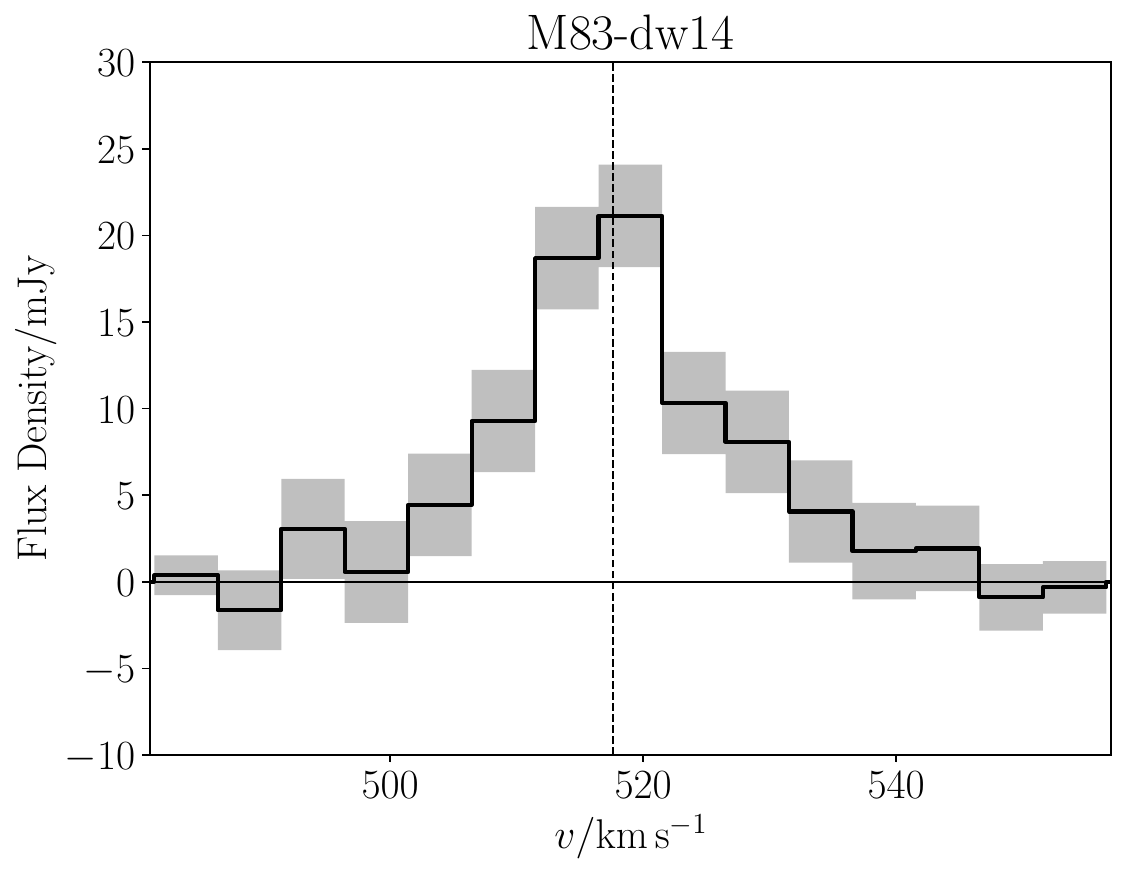}
    \caption{\textit{Top row}: VLA \hi \ contour overlays on DECaLS $griz$ images of Ursa Major C (left), Hydra B (center) and M83-dw14 (center). The contour levels begin at 3$\sigma$, and each subsequent contour is double the previous. The associated column density levels are given in Table~\ref{tab:HIobs}. The gray ellipse in the bottom right corner indicates the size and orientation of the beam. 
    \textit{Bottom row}: VLA \hi \ line profiles of the same sources extracted from the masked data cubes. The vertical dashed lines show the central velocities and the grey bands show the 1$\sigma$ uncertainties in the flux densities of each 5~\kms \ channel.}
    \label{fig:HIoverlays}
\end{figure*}

The VLA \hi \ data cubes were masked using the Source Finding in Astronomy tool \citep[SoFiA~v2;][]{SoFiA,SoFiA2}. We used the smooth and clip algorithm to perform masking. The parameters were fine-tuned for each source, but generally consisted of spatial smoothing between half and a full beam width, and velocity smoothing over 15~\kms, with the clipping threshold set to 4$\sigma$. In all cases SoFiA's reliability threshold was set to 99\%.

The flux-weighted central velocities, the velocity widths, and the fluxes calculated by SoFiA are shown in Table~\ref{tab:dwarfs}, along with their uncertainty estimates, for Ursa Major~C, Hydra~B, and M83-dw14. The corresponding masked spectral profile and moment zero maps are shown in Figure~\ref{fig:HIoverlays}. Cen~C, which was within the primary beam of the VLA observations of M83-dw14, was undetected. We extracted a spectrum centered on its position (see Appendix~\ref{sec:CenC_spec}) and estimated a 3$\sigma$ upper limit on its \hi \ flux for an assumed velocity width of 25~\kms \ \citep[cf. Leo~P;][]{Giovanelli13}. This \hi \ flux limit and the fluxes of the detections were converted to \hi \ masses using the standard conversion, $M_{\rm HI} = 2.356\times10^5 D^2 S_{\rm HI}\ M_\odot$ from \citet{Haynes98}, where $D$ is distance in Mpc and $S_{\rm HI}$ is the total flux in Jy~\kms.

The low Declinations of Hydra~B and M83-dw14 mean that the synthesized beam in both observations is considerably elongated in the north-south direction (Figure~\ref{fig:HIoverlays}), while that of Ursa Major~C is close to circular. In addition, as M83-dw14 was observed in C-configuration, the synthesized beam is smaller than for the other two detections. Both Hydra~B and M83-dw14 appear to be at best marginally resolved spatially, and relatively little can be inferred about their gas morphology. The elongation of M83-dw14's gas distribution is almost certainly driven by the beam shape, while the apparent low S/N extension to the north of Hydra~B is likely the result of low-level residual RFI (which heavily impacted these observations). 

However, the morphology of Ursa Major~C is highly irregular and clearly spatially extended even at D-configuration angular resolution. We were initially suspicious of this unusual morphology and re-reduced the data to fine tune the RFI excision. However, there was minimal RFI in the data to begin with and the resulting map was qualitatively unchanged. Given the low velocity of the \hi, we also considered the possibility that this gas cloud could be a chance superposition. Although a redshift of the optical component would be the only means of conclusively refuting this, this region of the sky appears to have minimal high velocity cloud (HVC) structure in the vicinity of this target \citep{Westmeier2018}. The closest HVCs that we identified in the literature were CHVC101.99+60.39+183 and CHVC101.79+60.28+193 \citep{Xu2026}, which are both just under 5$^\circ$ away and at $\sim$190~\kms.

Taking the morphology at face value, the gas column density is considerably lower than the other two detections, both of which have their peak column densities centered on the stellar body. In the case of Ursa Major~C, the lower column density gas is both offset from the stellar body and considerably extended to the south. There is no apparent companion to Ursa Major~C (see Section~\ref{sec:env}) that could have caused this irregular morphology through gravitational interaction, suggesting that the morphology is the result of internal processes. We discuss the interpretation of this further in Section~\ref{sec:discussion}.

Ursa Major~C's \hi \ content also, remarkably, does not fall on the \hi \ size--mass relation, which almost every known gas-bearing galaxy obeys \citep{Wang2016,Stevens2019}. Technically Ursa Major~C has no \hi \ diameter (as it is typically measured), because its peak column density (after VLA D-configuration beam smearing) is only $2.2 \times 10^{19}\;\mathrm{cm}^{-2}$, and \hi \ diameters are usually measured at a level of $1\;M_\odot\,\mathrm{pc}^{-2} \approx 1.25\times 10^{20}\;\mathrm{cm}^{-2}$. However, using the relation to convert Ursa Major~C's \hi \ mass to an expected diameter gives approximately 0.4~kpc, which is many times smaller than the extent of the \hi \ distribution visible in Figure~\ref{fig:HIoverlays} (regardless of beam smearing). Given the near total universality of the \hi \ size--mass relation, even for considerably disturbed morphologies, we expect that in higher resolution imaging Ursa Major~C's \hi \ distribution would likely break up into multiple clumps, each of which would be more compact and higher column density, and could obey the relation. 

We also inspected the moment one velocity maps of the three targets with VLA imaging. Ursa Major~C's velocity field does not show consistent structure, as might be expected, and Hydra~B is essentially entirely unresolved. M83-dw14 does show a weak velocity gradient that may be roughly aligned which the major axis of its stellar body. This could be an indication that it is rotationally supported, however, higher angular resolution data would be need to verify this.

Finally, Grapes was observed with the GBT rather than the VLA, and its \hi\ properties are measured directly from the calibrated GBT spectrum (see Table~\ref{tab:dwarfs}).

\begin{figure*}
    \centering
    \includegraphics[width=0.2\linewidth]{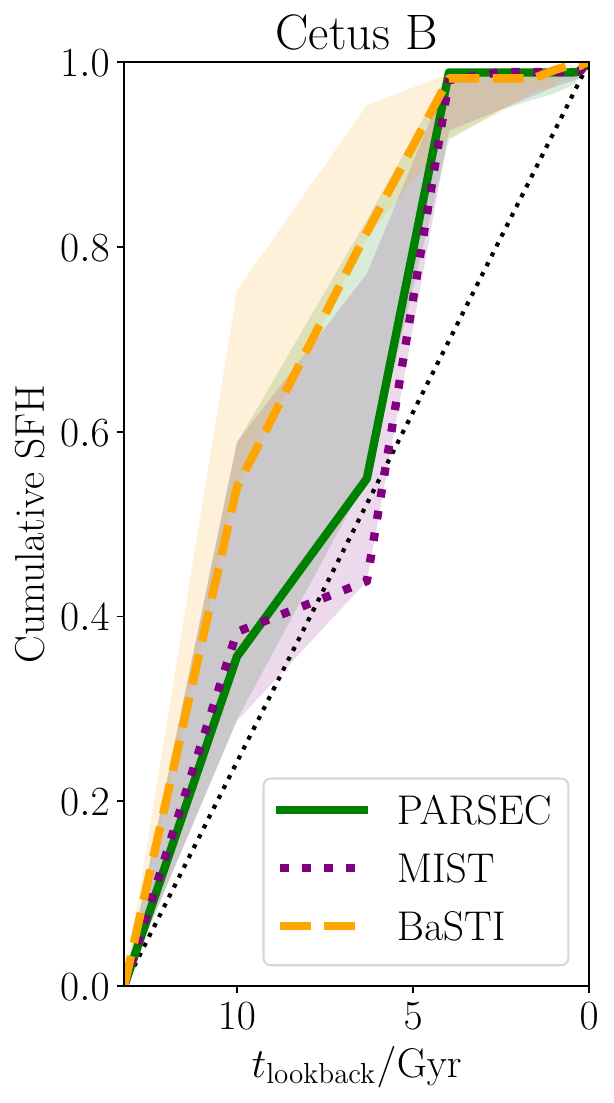}
    \includegraphics[width=0.2\linewidth]{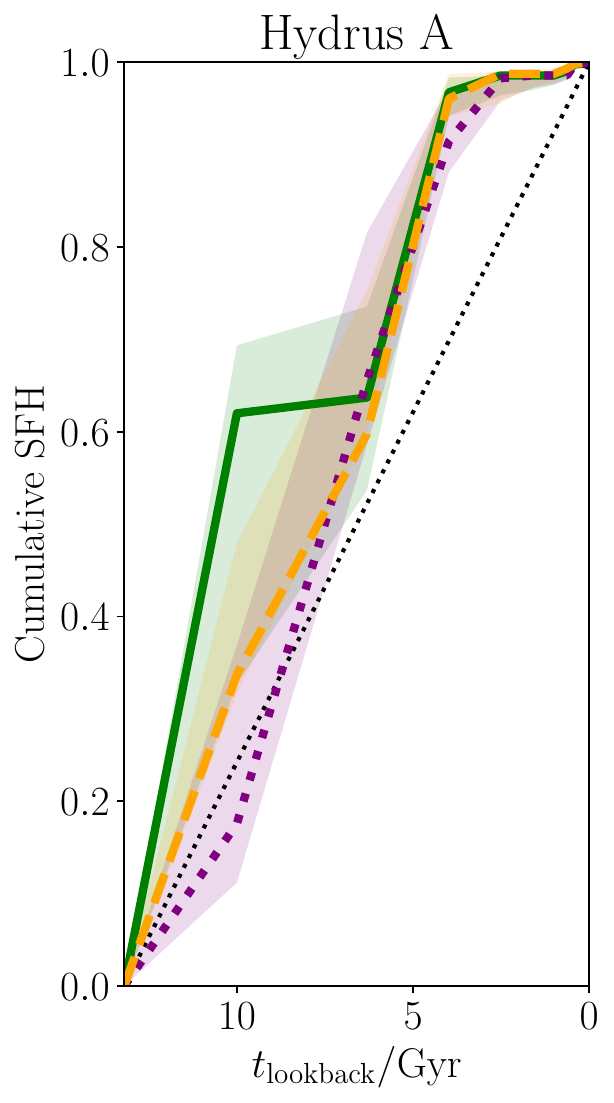}
    \includegraphics[width=0.2\linewidth]{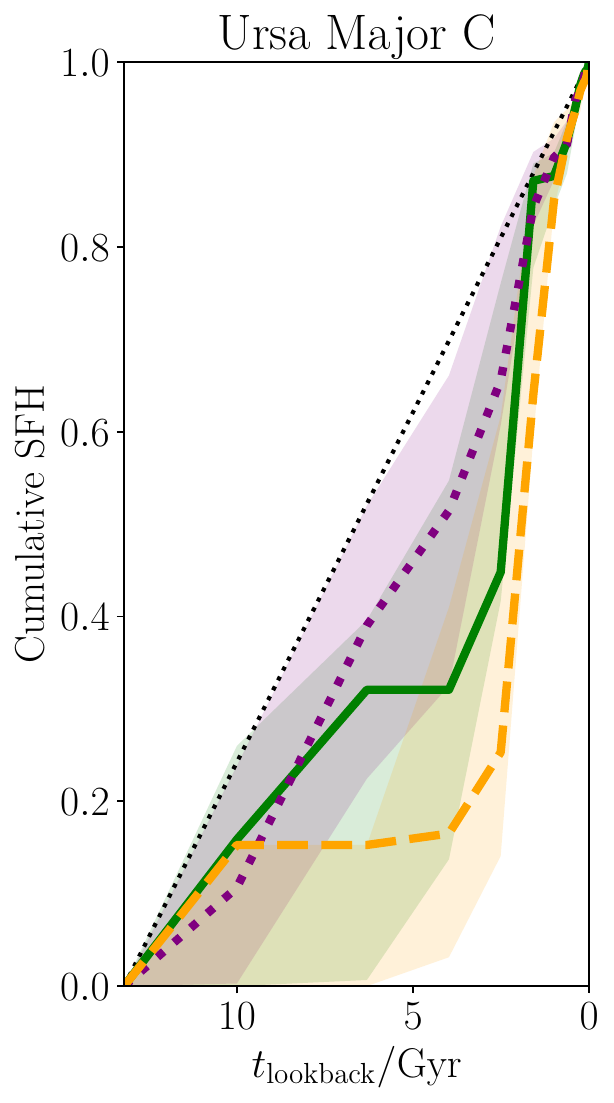}
    \includegraphics[width=0.2\linewidth]{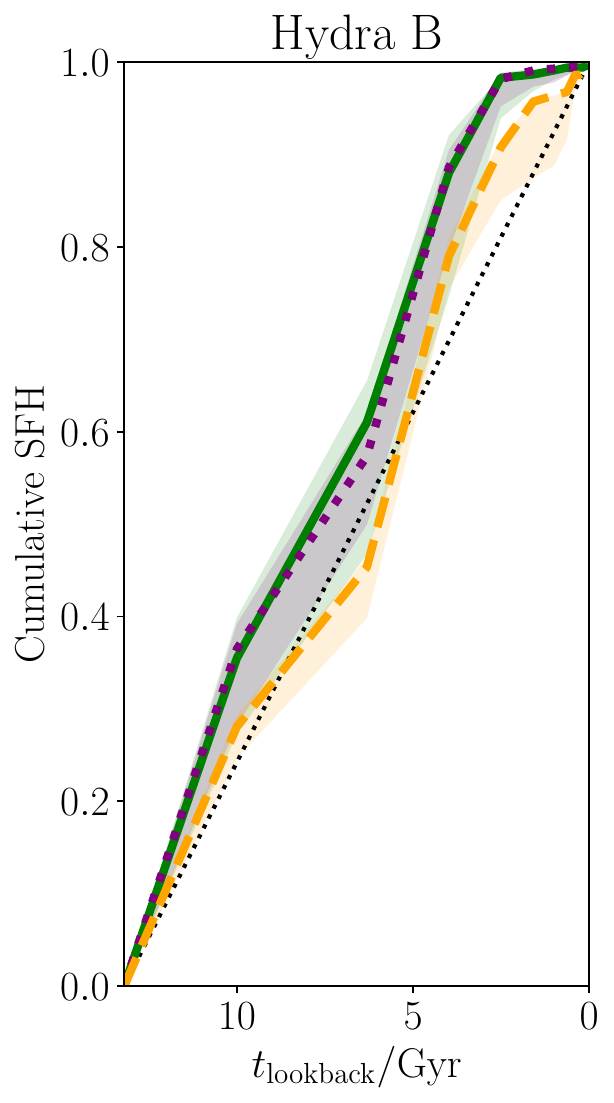}
    \includegraphics[width=0.2\linewidth]{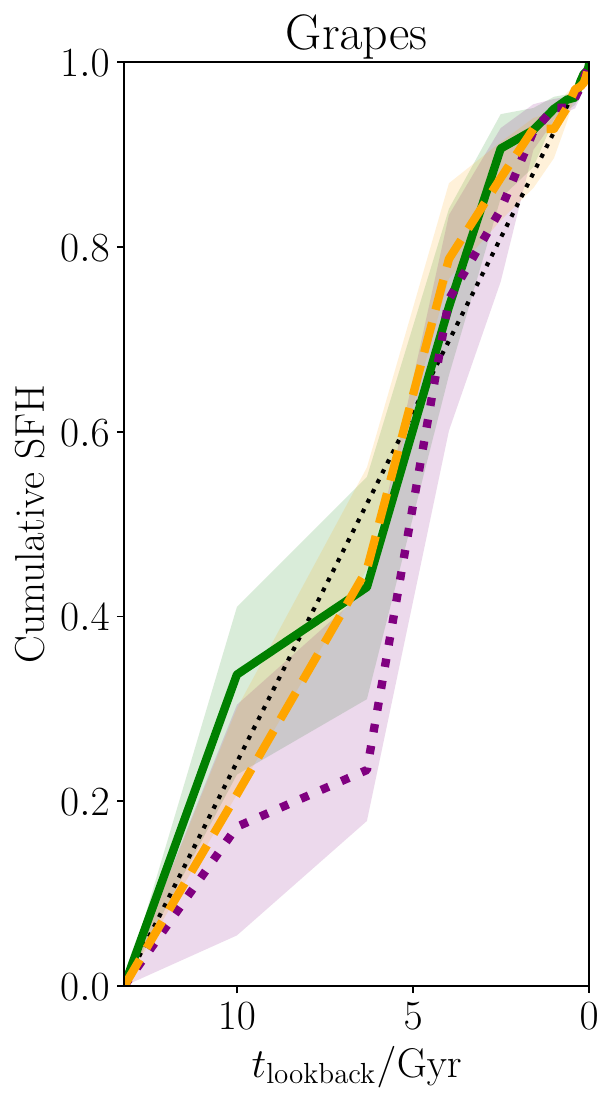}
    \includegraphics[width=0.2\linewidth]{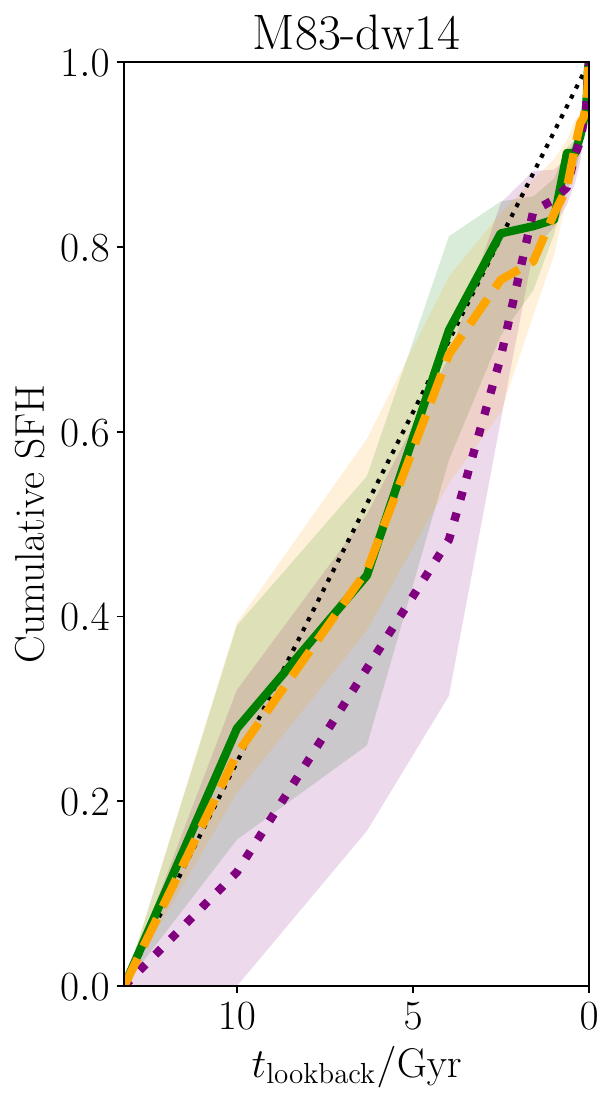}
    \includegraphics[width=0.2\linewidth]{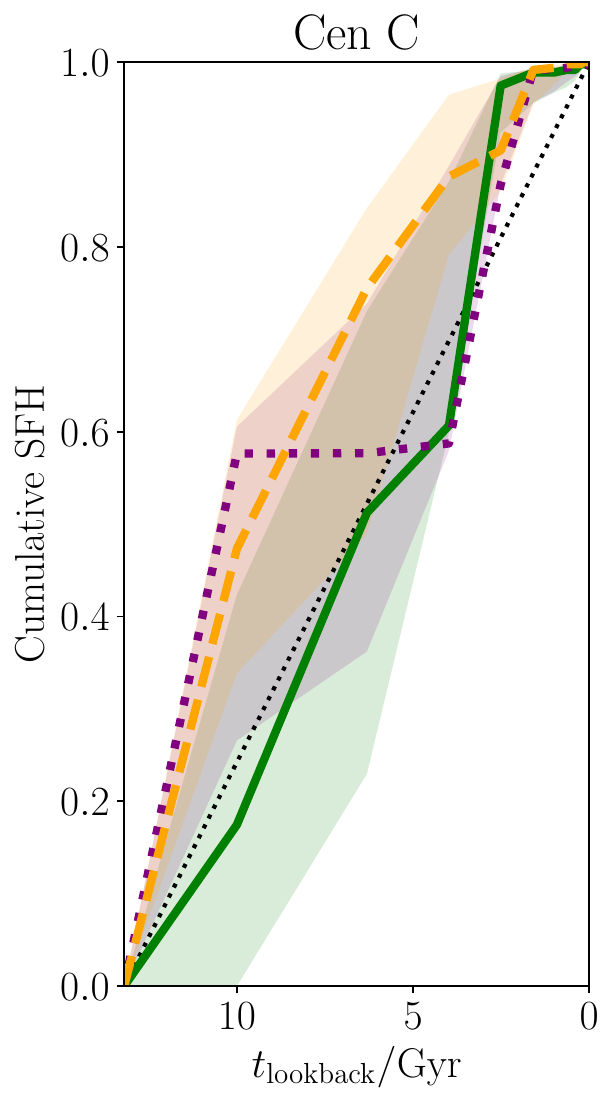}
    \caption{Cumulative star formation histories of our targets, showing the fraction of the total stellar mass formed as a function of lookback time. The fits from the three different isochrone libraries are shown with different colors and line styles. The diagonal dashed black lines represent a constant SFR for the age of the universe. The shaded bands indicate 1$\sigma$ uncertainties for each isochrone library separately. Note that the uncertainties for Hydra~B are likely underestimated (see discussion in text).}
    \label{fig:SFHs}
\end{figure*}

\subsection{Star Formation Histories \label{sec:sfh}}

We derive star formation histories (SFHs) using the same CMD-fitting methodology described in \citet{Mike25_sfh}, which follows the StarFISH approach \citep{Harris+2001}. In brief, observed CMDs are modeled as linear combinations of artificial single stellar population CMDs spanning a range of ages and metallicities. Synthetic CMDs are constructed using multiple isochrone libraries -- PARSEC \citep{Bressan2012}, MIST (MESA Isochrones and Stellar Tracks; \citealt{Dotter2016,Choi2016}), and BaSTI (the Bag of Stellar Tracks and Isochrones; \citealt{Pietrinferni2004,Hidalgo2018}) -- and incorporate completeness and photometric uncertainties derived from artificial star tests. The models span logarithmic ages of $\log(t/\mathrm{yr})=6.5$--$10.1$ and low metallicities, ranging from [M/H] = $-2.2$ to $-1.3$ for PARSEC and from $-2.5$ to $-1.3$ for MIST and BaSTI. The only exception is Hydra~B, which is slightly more massive than the other galaxies in our sample, for which we extend the upper metallicity limit to [M/H] = $-1.0$ for all three models. We allow metallicity to vary with age, imposing only weak constraints on the permitted age--metallicity combinations to avoid unphysical stellar populations.
A maximum likelihood fit between the observed (binned) CMD and the single stellar populations assigns weights to all populations, with the weights directly corresponding to the best-fit SFH. We refer the reader to \citet{Mike25_sfh} for full details of the implementation and uncertainty estimation.

Figure~\ref{fig:SFHs} presents the cumulative SFHs. Overall, the galaxies exhibit diverse evolutionary histories, ranging from systems dominated by early star formation to those with more extended or delayed stellar mass assembly. Several dwarfs, including Ursa Major~C and M83-dw14, show evidence for prolonged star formation over a large fraction of cosmic time, while Cetus~B, Hydrus~A, and, perhaps, Cen~C appear to have formed a substantial fraction of their stellar mass at early epochs, and are now quenched. We note that the earliest time bins should be treated with caution, as the depth of the current photometry allows significant degeneracies among ancient simple stellar populations, limiting our ability to robustly distinguish detailed star formation activity at the oldest epochs. Despite noticeable differences among the stellar evolution models, particularly at ancient lookback times ($\gtrsim 8$~Gyr), the broad evolutionary trends remain consistent across all libraries. In particular, Cetus~B, Hydrus~A, and Cen~C show little to no evidence for recent star formation, 
as expected from the absence of young stellar populations in the CMDs shown in Figure~\ref{fig:cmd}. In contrast, others exhibit more recent star formation activity, consistent with the presence of young stellar populations seen in Figure~\ref{fig:cmd2}.

The SFHs are also consistent with the gas content of these systems. Cetus~B, Hydrus~A, and Cen~C are all undetected in \hi, as expected given their lack of recent star formation. In contrast, the \hi-detected galaxies (Ursa Major~C, Hydra~B, Grapes, and M83-dw14) all show evidence of recent star formation, although at varying levels. Overall, the presence of \hi\ tracks recent star formation activity, and we find good agreement between the inferred SFHs and the observed gas properties.

\subsection{Environment \label{sec:env}}

\begin{figure*}[!ht]
\centering
\includegraphics[width = 0.45\linewidth]{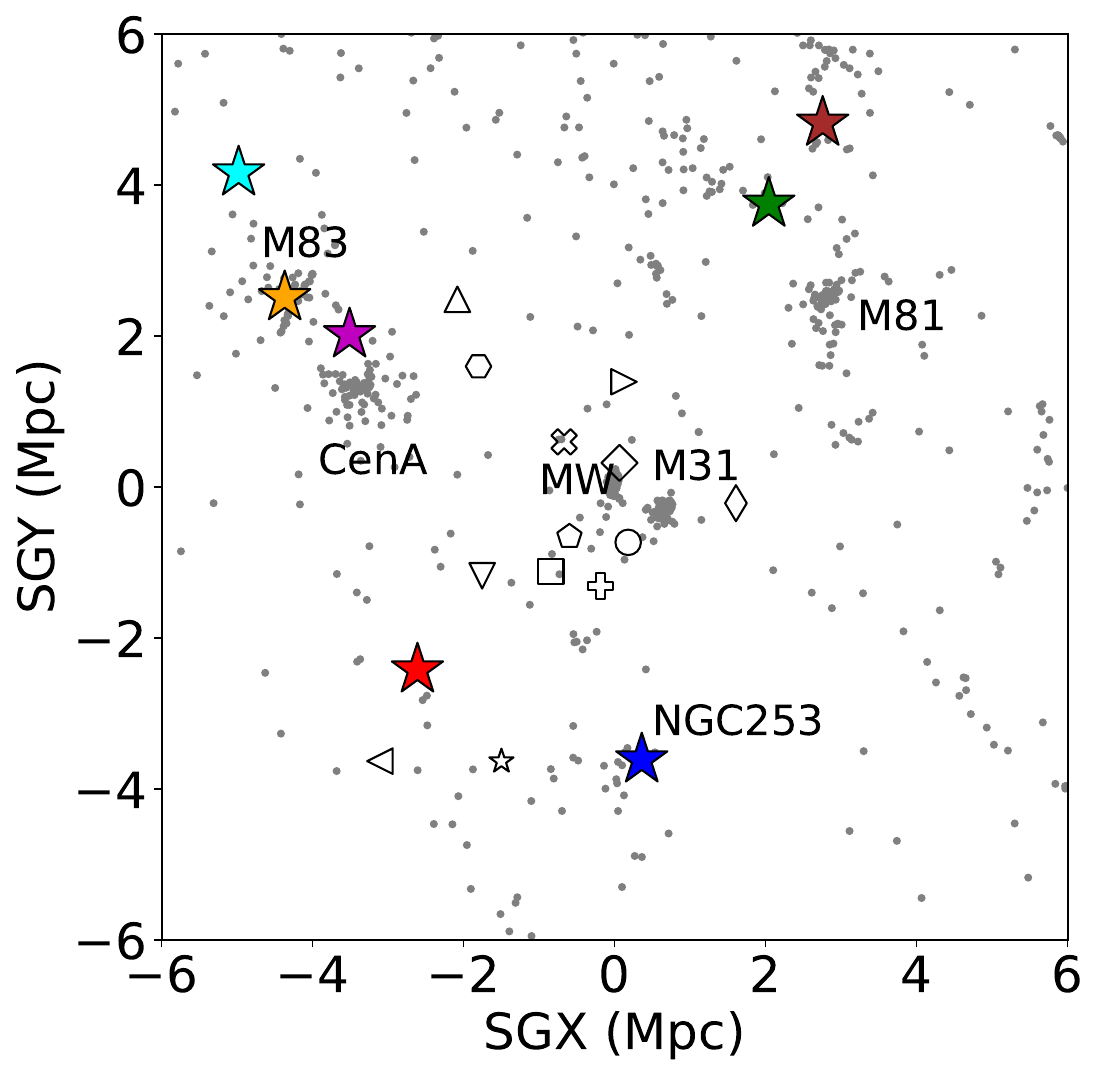}
\includegraphics[width = 0.45\linewidth]{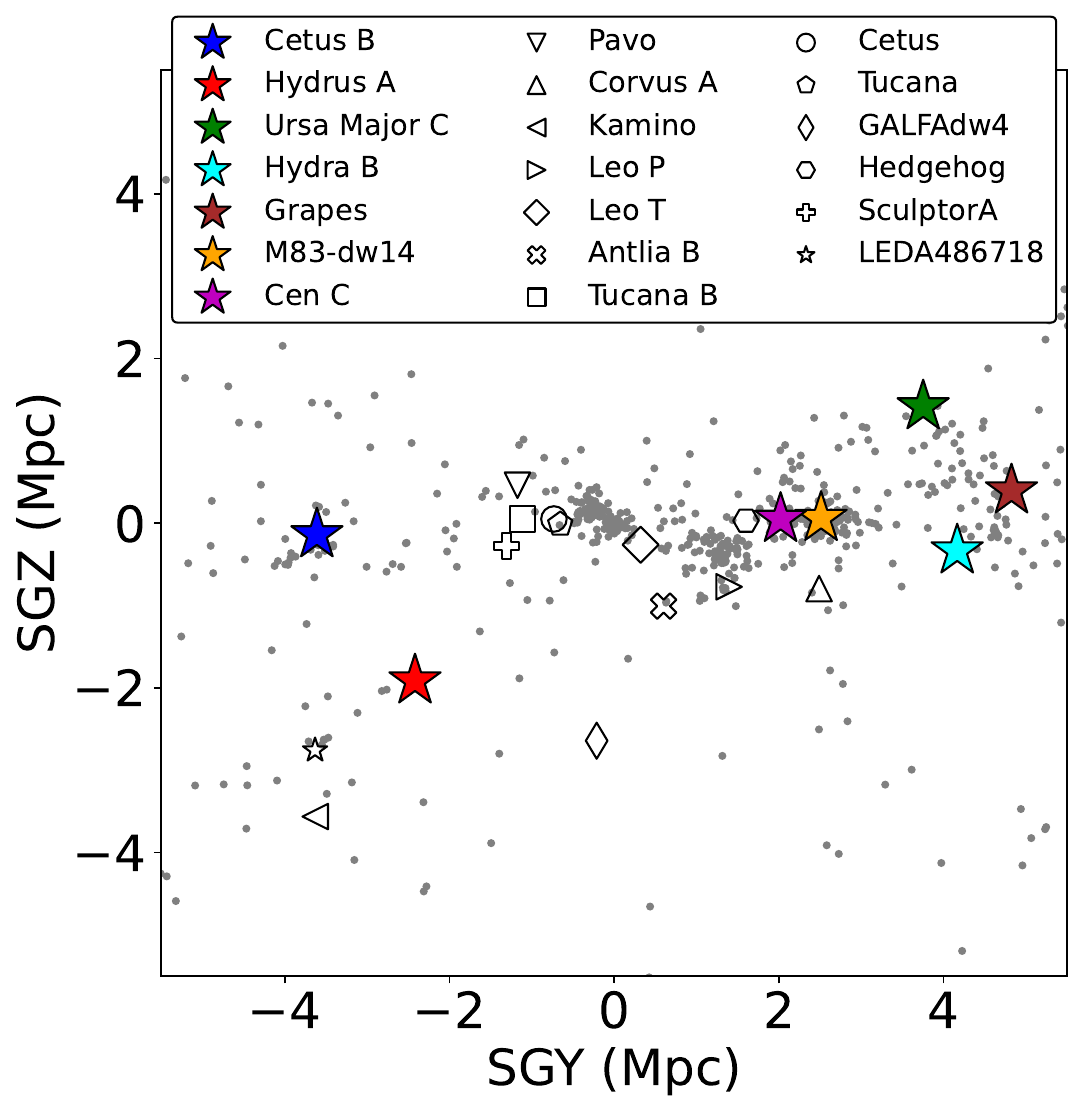}
\caption{A map of nearby galaxies in supergalactic coordinates, illustrating the environment of our targets. Grey points are taken from the compilation of \citet{Karachentsev2019}. Left: The supergalactic $X-Y$ plane. Right: We show the supergalactic $Y-Z$ plane and the local plane of galaxies.  
Cetus~B, Hydrus~A, Ursa Major~C, Hydra~B, M83-dw14, and Cen~C are marked with stars. Other notable nearby dwarf galaxies are included for comparison. 
\label{fig:env}}
\end{figure*}

\begin{figure*}[!hb]
\centering
\includegraphics[width = 0.42\linewidth]{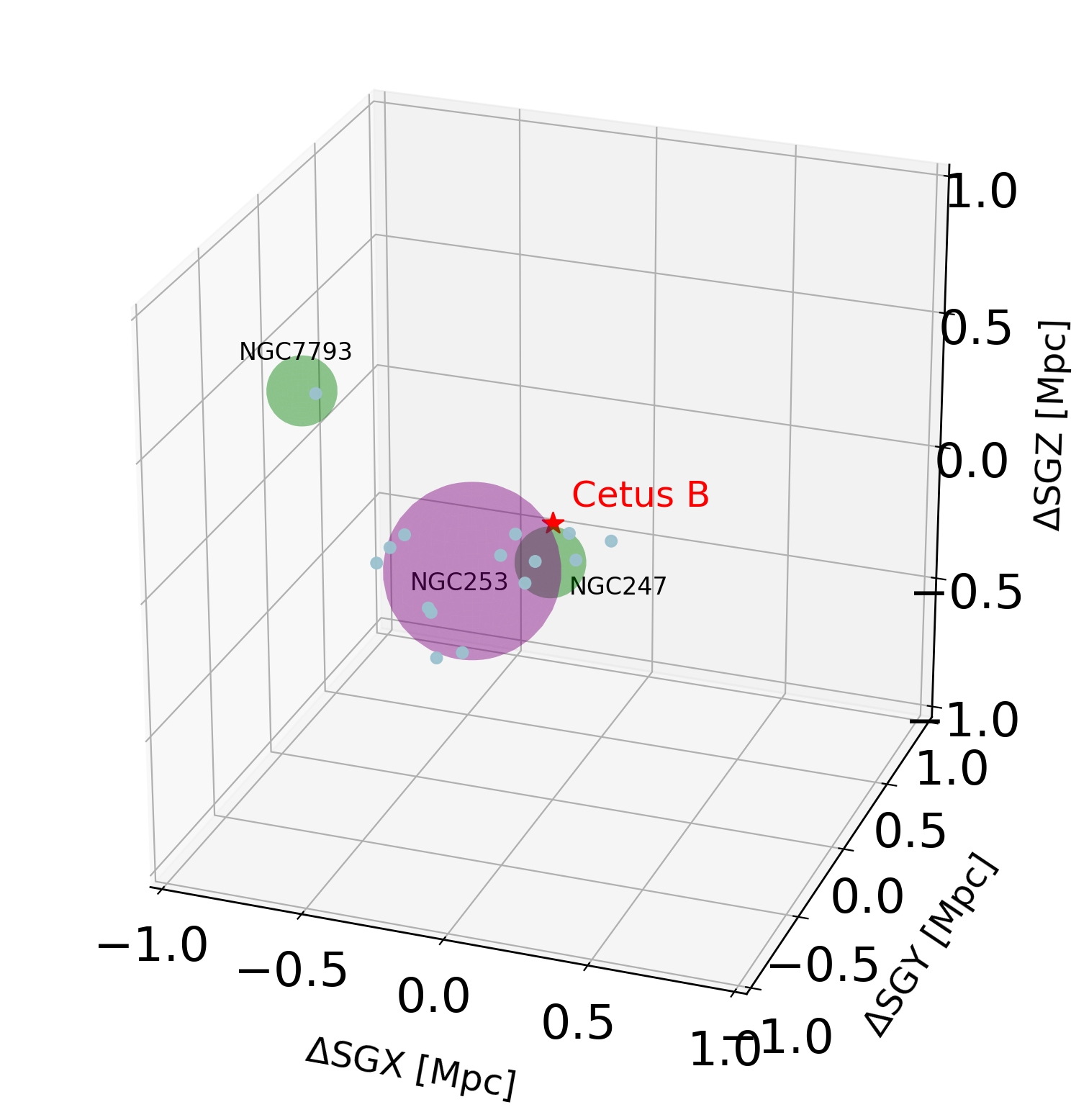}
\includegraphics[width = 0.42\linewidth]{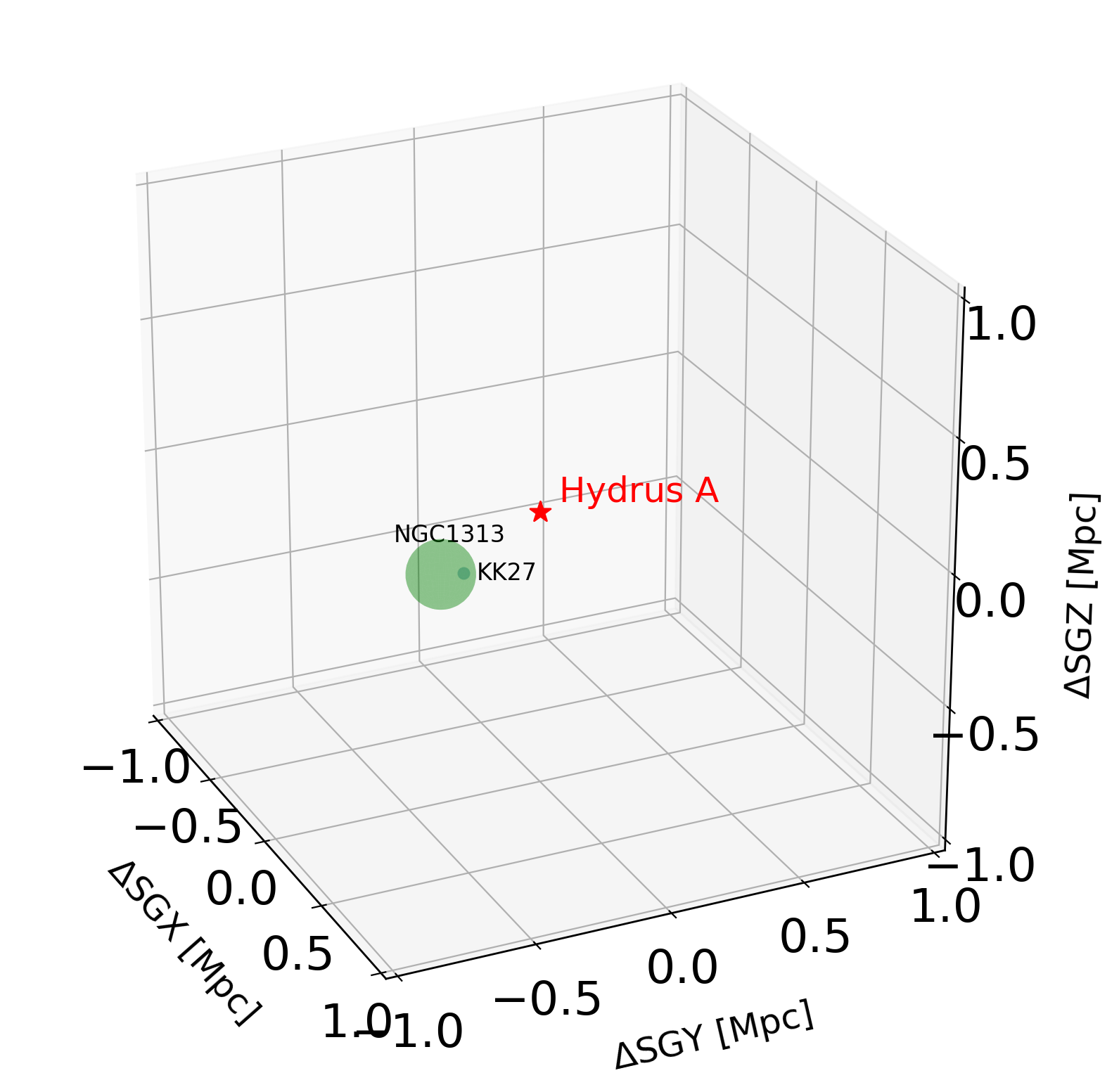}
\includegraphics[width = 0.42\linewidth]{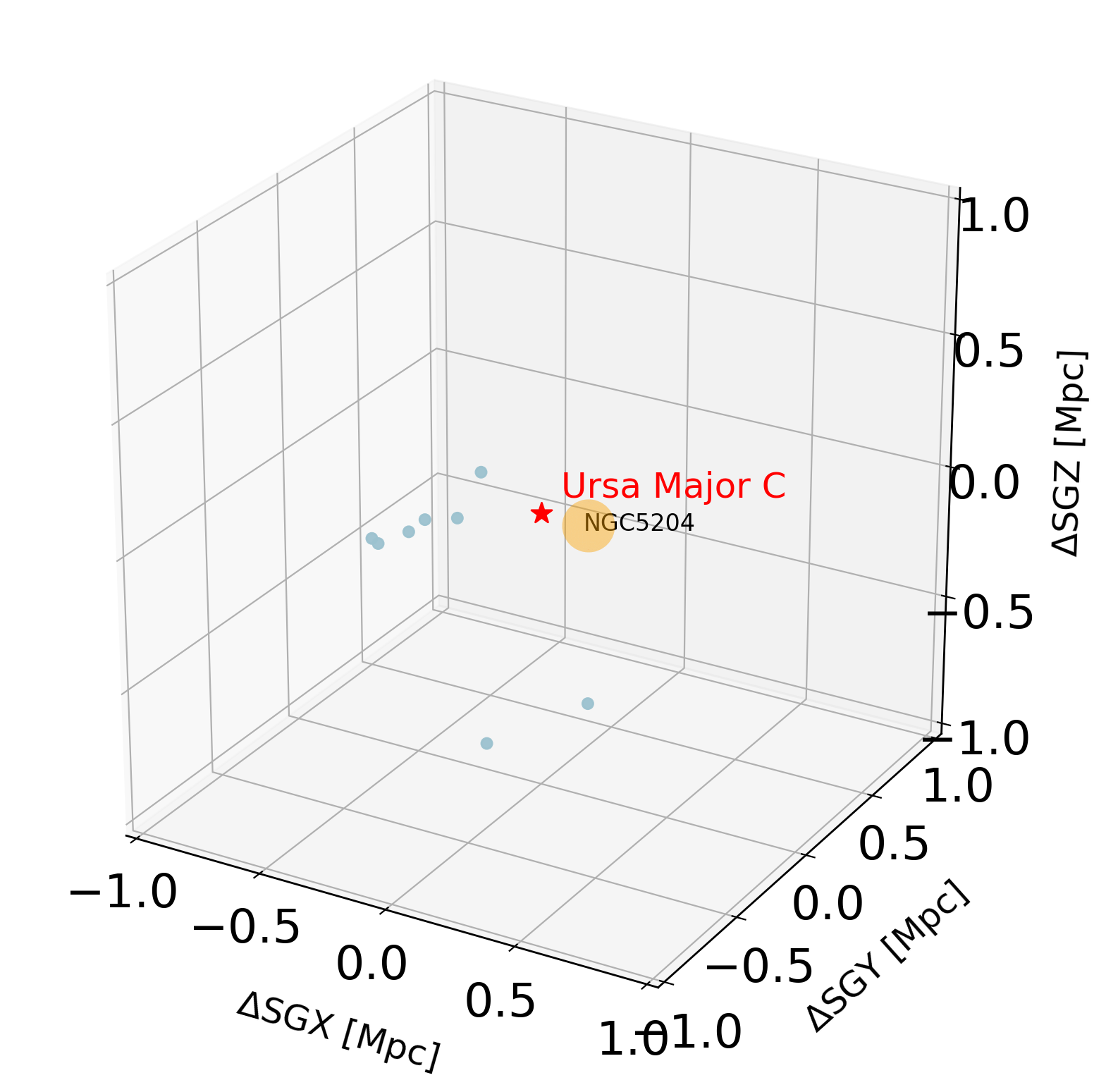}
\includegraphics[width = 0.42\linewidth]{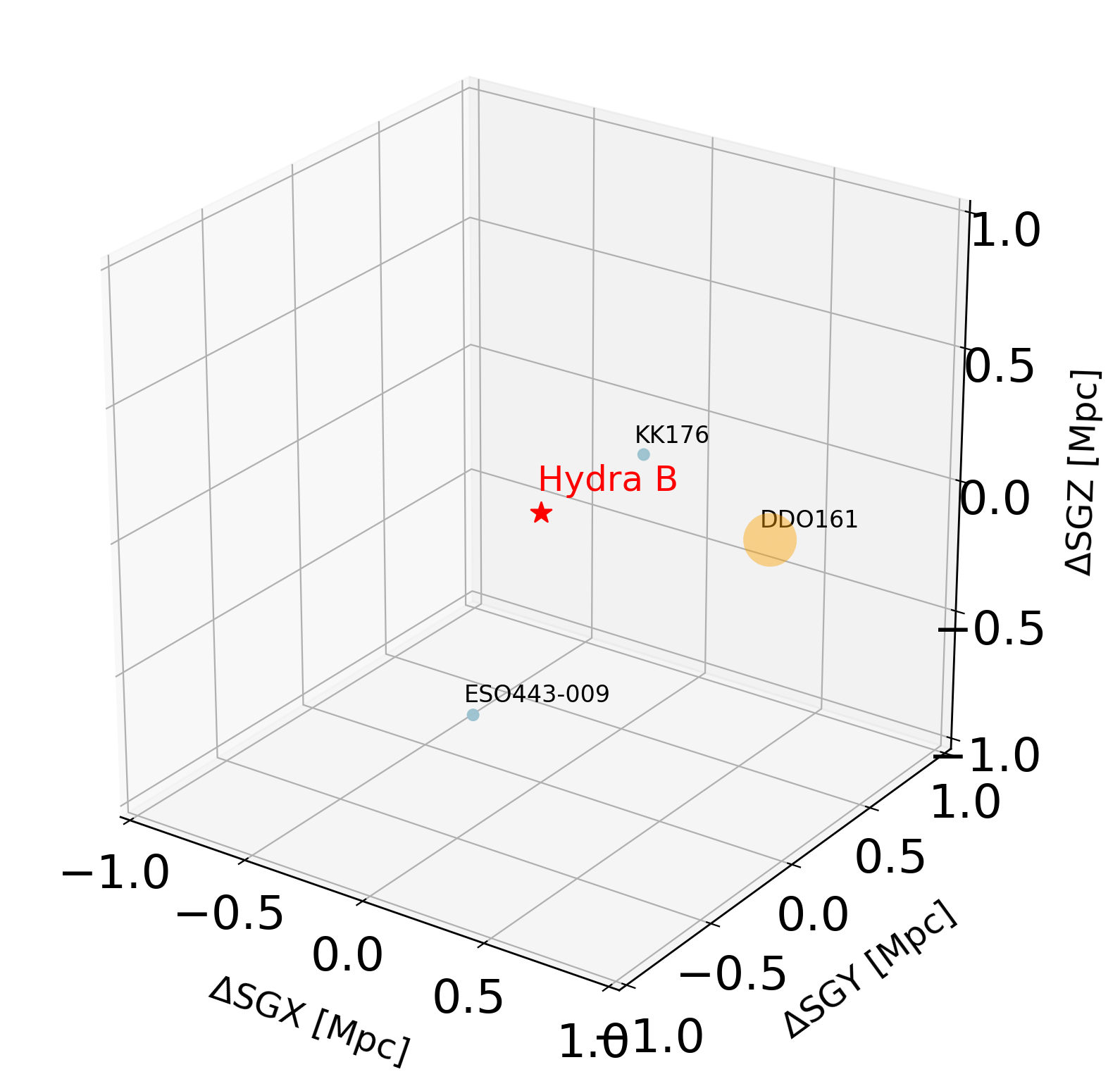}
\includegraphics[width = 0.42\linewidth]{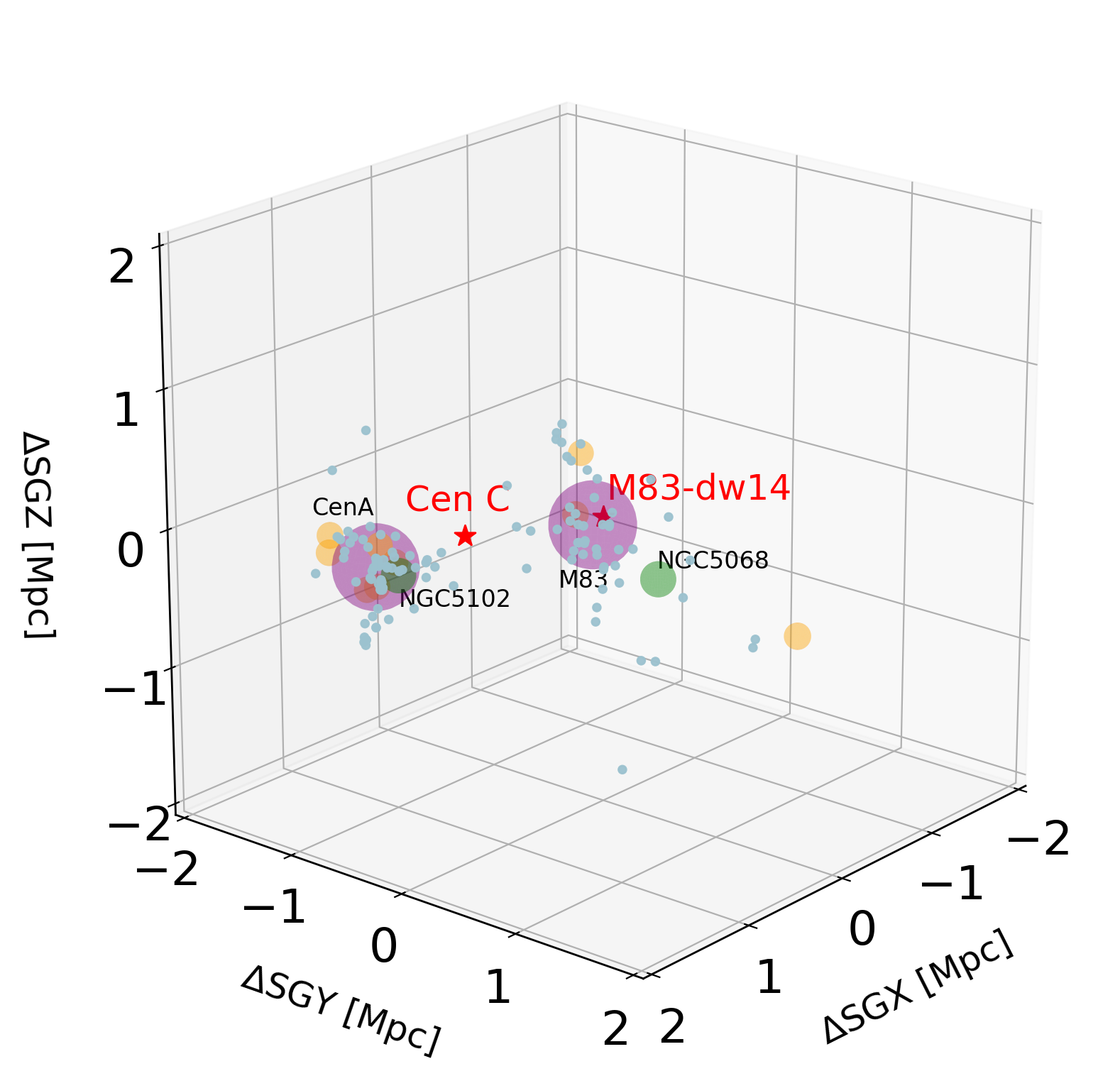}
\includegraphics[width = 0.42\linewidth]{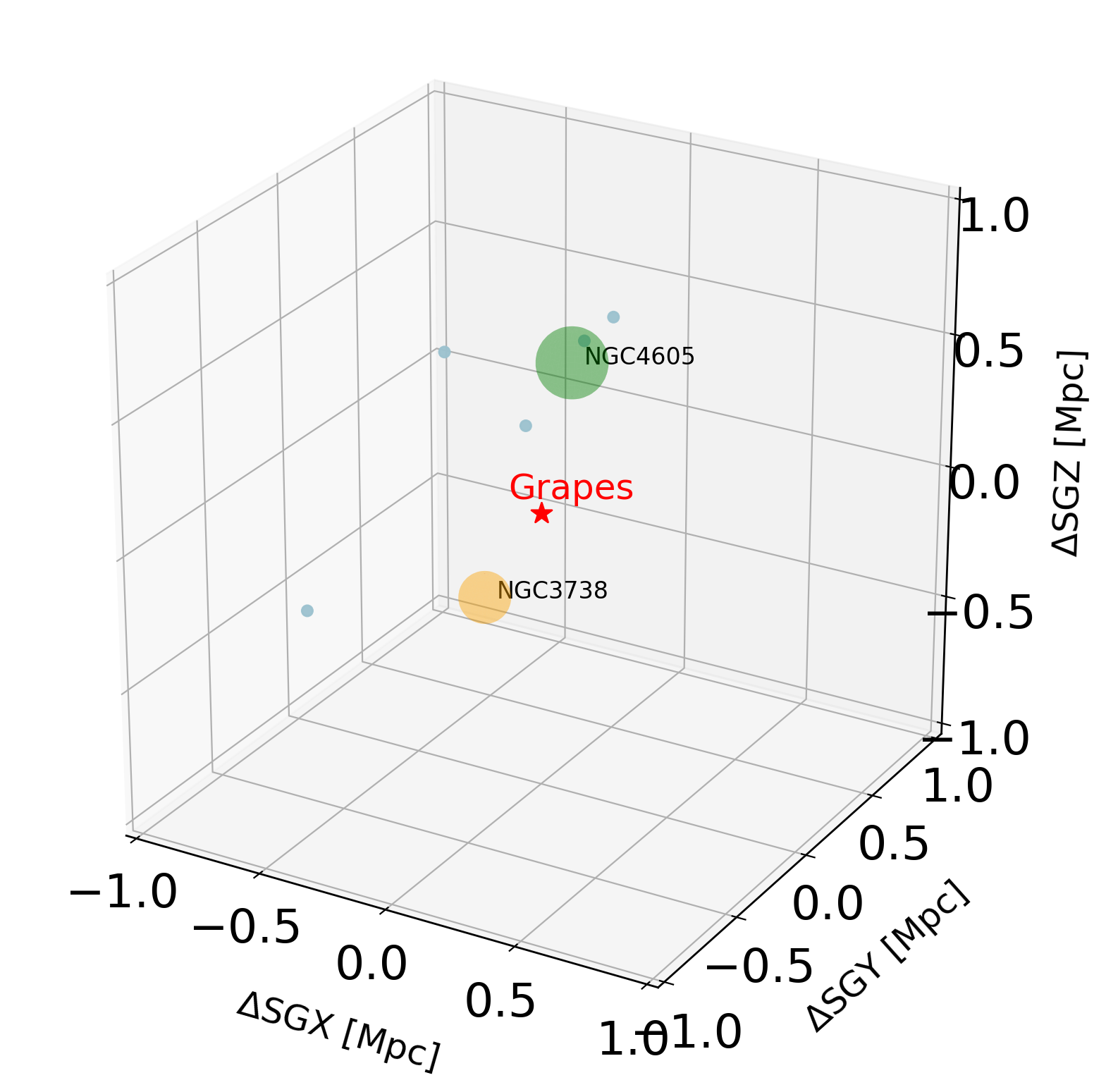}
\caption{A close-up view of the environment of each dwarf in 3D. The purple, green, and orange spheres represent the approximate virial radii of MW-mass, LMC-mass, and SMC-mass galaxies, respectively (see text for details), while light-blue points indicate lower-mass dwarfs. Our targets span environments ranging from likely satellites of massive halos to relative isolation. 
\label{fig:env2}}
\end{figure*}

To explore the environment of our targets, we plot their positions, together with other nearby galaxies within 7.5~Mpc from \citealt{Karachentsev2019}, in two projections of the supergalactic coordinate system in Figure~\ref{fig:env}.  
We also highlight several other dwarfs in low-density or isolated environments for comparison. Cetus~B lies in projected proximity to the NGC~253 group, while Hydra~B, M83-dw14, and Cen~C are projected near the M83/Cen~A group. Ursa Major~C and Grapes lie within the local sheet, but in directions away from any known nearby group. Hydrus A, by contrast, is notably distant from the local sheet or any known major group.

In Figure~\ref{fig:env2}, we show the three-dimensional (3D) locations of our targets relative to known nearby galaxies. We divide the known galaxies from \citet{Karachentsev2019} into four categories based on stellar mass. For the three highest-mass categories, we represent each galaxy by a colored sphere whose radius approximates the expected virial radius \citep[e.g.,][]{BMP2021}. Galaxies with $M_{\star}$$\geq$$10^{10} M_{\odot}$, corresponding to MW-mass or more massive systems, are shown as purple spheres with assumed virial radii of 300~kpc. Galaxies with $10^9$$\leq$$M_{\star}/M_{\odot}$$<$$10^{10}$, corresponding approximately to LMC-mass systems, are shown as green spheres with assumed virial radii of 120~kpc. Galaxies with $10^8$$\leq$$M_{\star}/M_{\odot}$$<$$10^{9}$, corresponding approximately to SMC-mass systems, are shown as orange spheres with assumed virial radii of 90~kpc. Lower-mass dwarfs with $M_{\star}/M_{\odot}$$<$$10^{8}$ are shown as light-blue points. Stellar masses are estimated from the $K_s$-band luminosities by adopting a mass-to-light ratio of $M_{\star}/L_{K_s}=0.6$ \citep{McGaugh2014}.

Our targets span environments ranging from association with nearby massive halos to relative isolation, providing a useful baseline for examining how environment relates to the structural and star formation properties derived from our HST imaging. Cetus~B and M83-dw14 are associated with nearby massive galaxies. Cetus~B lies near NGC~253 and NGC~247, at 3D separations of 327 and 134~kpc, respectively. These separations place Cetus~B just outside the adopted virial radii of both galaxies. However, its present-day location does not preclude a past interaction. Cetus~B could be a backsplash galaxy that previously passed within the virial radius of either galaxy, or it could be on its first infall into the broader NGC~253/NGC~247 group environment. M83-dw14, by contrast, lies within the adopted virial radius of M83, at a 3D separation of 236~kpc. Its \hi\ systemic velocity is also similar to that of M83 \citep{Meyer2004}, providing additional kinematic support for its classification as a satellite.

Cen~C is also located within the broader M83/Cen~A complex, although it is not directly associated with the halo of either massive galaxy. Despite being discovered serendipitously in the same HST field as M83-dw14, Cen~C is separated from it by 990~kpc in 3D and is therefore not physically associated with M83-dw14. Cen~C lies approximately midway between M83 and Cen~A, at 3D separations of 867 and 848~kpc, respectively. Its nearest known neighbor is IC~4316, at a 3D separation of 377~kpc. Thus, Cen~C occupies a relatively isolated location within the broader M83/Cen~A complex.

The remaining systems reside in relatively low-density environments. Hydrus~A is separated by 428~kpc in 3D from the LMC-mass galaxy NGC~1313, while its nearest known neighbor, KK~27, lies 381~kpc away. The nearest known neighbor of Ursa Major~C is the SMC-mass galaxy NGC~5204, at a 3D separation of 181~kpc, approximately twice its adopted virial radius. Hydra~B is even more isolated: its nearest relatively massive galaxy, the SMC-mass system DDO~161, is 834~kpc away in 3D, and its nearest known neighbor, ESO~443-009, lies 827~kpc away. Grapes lies between the LMC-mass galaxy NGC~4605 and the SMC-mass galaxy NGC~3738, at 3D separations of 898 and 351~kpc, respectively. Although NGC~3738 is its nearest known neighbor, Grapes lies well beyond its adopted virial radius ($\sim90$~kpc).

\section{Discussion and Conclusions}\label{sec:discussion}

\begin{figure}[!t]
\centering
\includegraphics[width = \linewidth]{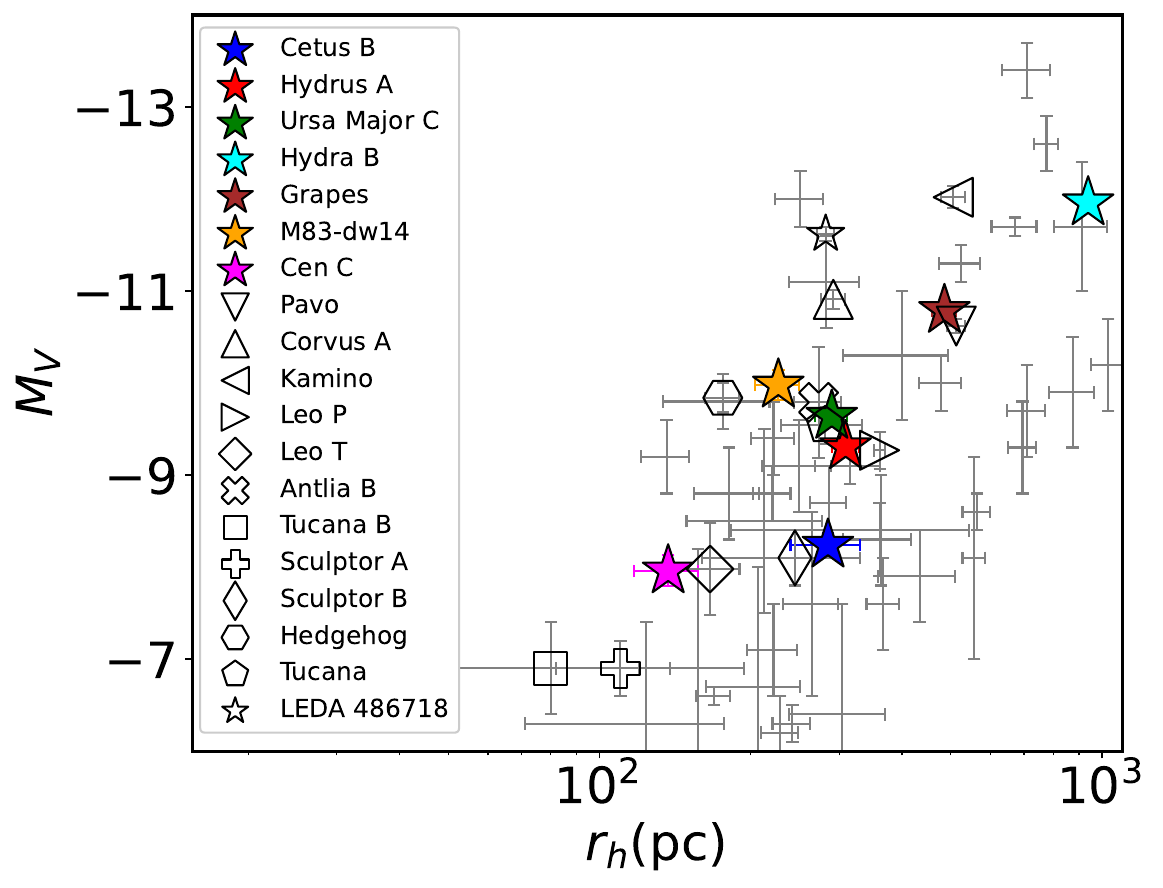}
\caption{Absolute magnitude as a function of half-light radius of our targets, relative to Local Volume dwarfs \citep{McConnachie2012,McQuinn2024,Bennet2022,Sand2022,Li2024,Sand2024}. Plotted as grey points are the Local Group dwarf satellites for context \citep{LVDB2024}. For some data points, the error bars are smaller than the symbols and are therefore not visible.
\label{fig:mv-rh}}
\end{figure}

We present HST imaging of seven Local Volume dwarf galaxies from the SEAMLESS survey: Cetus~B, Hydrus~A, Ursa Major~C, Hydra~B, Grapes, M83-dw14, and our serendipitous discovery, Cen~C. The imaging is new for all dwarfs except Grapes, for which we use archival HST data. Our data resolve each dwarf into stars, allowing us to measure their distances, structural properties, recent SFHs, and environments. In Figure~\ref{fig:mv-rh}, we compare our dwarfs with Local Group dwarfs and several other notable dwarfs within the Local Volume in the size-luminosity parameter space. All seven systems fall within the locus occupied by known dwarfs, with absolute magnitudes $M_V$$\approx$$-7$ to $-12$ and half-light radii of $100-1000$~pc, consistent with faint dwarfs in these various environments.  However, compared to Local Group dwarfs of similar luminosities, our sample appears to preferentially occupy the more compact, higher-surface-brightness region of parameter space, with a relative lack of larger, more diffuse systems. This likely reflects a surface brightness selection bias in the SEAMLESS detection and/or follow-up confirmation process. Our ongoing survey completeness analysis, together with our upcoming larger scale follow-up programs using HST (SNAP-18410, PI: Jensen) and Roman (Cycle 1, ID: 19014, PI: Mutlu-Pakdil), will allow us to better quantify this potential bias and assess whether the trend persists across a larger sample.

In the following subsections, we compare each of our targets with galaxies occupying similar regions of the size-luminosity plane to investigate how systems with comparable structural properties differ in their star formation histories, gas content, and environments. 

\subsection{Cen~C}
Cen~C is among the faintest isolated dwarf galaxies known, only slightly brighter than the two isolated ultra-faint dwarfs Tucana~B \citep{Sand2022} and Sculptor~A \citep{Sand2024}. Based on its size and luminosity, Cen~C is the closest analog of Leo~T, the prototypical isolated transition dwarf in the Local Group \citep{deJong2008, Weisz2012}. However, the two occupy very different environments. Leo~T lies only 420~kpc from the Milky Way and is on its first infall \citep{Rocha2012,Bennet2025}. Cen~C, by contrast, is relatively isolated within the broader M83/Cen~A complex. Deep HST photometry of Leo~T resolves young main-sequence and blue helium-burning stars, a broad RGB, a horizontal branch, and the main-sequence turnoff \citep{Weisz2012}. Cen~C shows no analogous bright blue population, although such stars would be detectable at its distance if present. Nevertheless, their cumulative SFHs appear qualitatively similar: both formed roughly half of their stellar mass by intermediate lookback times and continued forming stars gradually until relatively recently, rather than undergoing rapid early quenching. However, given the large uncertainties in Cen~C's SFH, this apparent similarity should be interpreted with caution, particularly as its CMD is dominated by old stars (see Figure~\ref{fig:cmd}). Leo~T also shows faint GALEX emission \citep{Lee2011}, whereas Cen~C has no clear GALEX counterpart, consistent with the absence of young blue stars in its CMD. Leo~T's extended star formation has been linked to its high baryonic gas fraction, $M_{\rm gas}/(M_{\star}+M_{\rm gas})\sim0.8$, with $\log(M_\mathrm{HI}/M_{\odot})=5.4$ \citep{RyanWeber2008}. In contrast, our non-detection of Cen~C places an upper limit of $\log(M_\mathrm{HI}/M_{\odot})<5.3$, meaning that an \hi\ reservoir comparable to that of Leo~T \citep[e.g.,][]{Adams2018} would have been detected in our observations. This is particularly intriguing given that Cen~C is more isolated than Leo~T, yet appears to be less gas-rich. Deeper \hi\ observations will be important for further constraining Cen~C's gas fraction and understanding what regulates gas retention and extended star formation in extremely low-mass isolated dwarfs.

\subsection{Cetus~B}
As first pointed out in \citet{Fielder2025}, Cetus~B closely resembles Sculptor~B, a quenched dwarf that may be a backsplash satellite of the LMC-mass galaxy NGC~300 \citep{Sand2024}. This comparison is particularly intriguing because Cetus~B may likewise be associated with an LMC-mass host, NGC~247. Just like Sculptor~B, Cetus~B has no clear GALEX counterpart, consistent with its quenched SFH. Together, these properties suggest that Cetus~B may have quenched during its current infall into the NGC~247/NGC~253 environment or during an earlier pericentric passage, in which case it would now be a backsplash galaxy.

\subsection{Hydrus~A}
Based on our updated structural measurements, Hydrus~A closely resembles Leo~P, the prototypical isolated, gas-rich, star-forming dwarf \citep{Giovanelli2013,McQuinn2015,McQuinn2024}. Unlike Leo~P, however, Hydrus~A shows no evidence of recent star formation: it has no GALEX UV counterpart \citep{Fielder2025}, indicating little or no activity within the past $\sim100$~Myr, consistent with its CMD and best-fit SFH. The current \hi\ limit for Hydrus~A ($\log(M_{\rm HI}/M_{\sun})<6.3$, \citealt{Fielder2025}) is too shallow to detect a reservoir comparable to that of Leo~P, $\log(M_{\rm HI}/M_{\sun})=5.91$ \citep{Giovanelli2013,McQuinn2024}, even if one were present. The two galaxies also occupy comparably isolated present-day environments. Hydrus~A lies 428~kpc from NGC~1313, while Leo~P is 412~kpc from Sextans~B. Despite this similarity, they exhibit different evolutionary states: Leo~P remains gas-rich and actively star-forming, whereas Hydrus~A appears quenched. This contrast may indicate that present-day isolation alone does not determine their evolution and leaves open the possibility of a past interaction between Hydrus~A and NGC~1313. Deeper \hi\ observations reaching Leo~P-like sensitivities are needed to determine whether Hydrus~A retains a similarly low-mass gas reservoir. Gas retention would argue against a past stripping event with NGC~1313 and instead favor reionization-suppressed accretion, whereas a deep non-detection would be consistent with an environmental stripping event. Such observations would help explain why two dwarfs so similar in size, luminosity, and present-day isolation have reached such different evolutionary states.

\subsection{Ursa Major~C}
Ursa Major~C occupies a size-luminosity regime shared by two very different systems: Antlia~B, a gas-rich but currently quenched satellite of NGC~3109 \citep{Sand15b,Hargis20}, and Tucana, an isolated dwarf spheroidal near the periphery of the Local Group \citep{Monelli2010,Savino2019}. These systems occupy distinct present-day environments. Antlia~B lies only $\sim75$~kpc from its SMC-mass host, and its quenching has been attributed to environmental suppression \citep{Hargis20}. Ursa Major~C lies 181~kpc from the similarly massive NGC~5204, or roughly twice its adopted virial radius. Although it is currently outside the halo boundary, a previous passage through the halo remains possible, in which case Ursa Major~C could be a backsplash system. Tucana is much more isolated today, at $\sim890$~kpc from the Local Group's massive galaxies, but may have interacted with the Milky Way until $\sim10$~Gyr ago \citep{Savino2019}. Such an encounter would likely have produced stronger environmental processing than an interaction between Ursa Major~C and NGC~5204. The SFHs of the three galaxies show broadly similar episodes of growth and reduced activity, but with different timing and outcomes. Antlia~B formed stars relatively steadily for its first $\sim10$-11~Gyr before becoming largely quiescent over the past $\sim2$--3~Gyr. Tucana experienced several early episodes of star formation, with its final episode ending $\sim6$-8~Gyr ago \citep{Savino2019}. Although the uncertainties are large, the SFH of Ursa Major~C similarly seems to have entered a lull $\sim7$-10~Gyr ago, but subsequently resumed star formation over the past $\sim4$~Gyr. This resurgence is consistent with the 75--500~Myr population visible in its CMD and distinguishes Ursa Major~C from both Antlia~B and Tucana. 

The \hi\ morphology of Ursa Major~C may provide an additional clue to the origin of this episodic behavior. Its \hi\ is strongly asymmetric, offset from the stellar body, and extended substantially to the south. Similar disturbed and spatially offset \hi\ has recently been observed in the extremely isolated dwarf Pavo, where the absence of plausible environmental perturbers and comparison with simulations point toward stellar feedback as the origin of the disturbed gas morphology \citep{Jones25}. Pavo and Ursa Major~C have almost identical \hi \ masses, but Pavo's stellar mass is about 0.4~dex greater than Ursa Major~C's. In such low-mass systems, supernova-driven outflows following episodes of star formation can temporarily displace and disrupt the neutral gas reservoir, suppressing subsequent star formation until gas re-accretes and the cycle begins again \citep{Rey2020,Rey2022}. The presence of a 75--500~Myr population in Ursa Major~C, together with its disturbed \hi\ morphology, raises the possibility that we are witnessing a similar feedback-driven cycle.  Higher resolution \hi\ observations would reveal more about the gas structure and kinematics of Ursa Major~C.

\subsection{M83-dw14}
Together with Antlia~B and Hedgehog, M83-dw14 spans nearly the full environmental range from confirmed satellite to extreme isolation, despite all three occupying a similar position in size and luminosity. M83-dw14 is a satellite of the MW-mass galaxy M83, while Antlia~B lies only $\sim75$~kpc from the substantially lower-mass NGC~3109 \citep{Hargis20}. Hedgehog is among the most isolated quenched dwarfs known, suggesting either quenching in isolation or a past backsplash passage through the Cen~A group \citep{Li2024}. Because no resolved SFH has yet been published for Hedgehog, the comparison is limited to its integrated properties and present-day quiescent state. Despite orbiting the most massive host of the three, M83-dw14 is also the most actively star-forming. Its CMD contains a clear blue sequence consistent with populations as young as 50-300~Myr, and its cumulative SFH rises steadily to recent times. It is also gas-rich, with over five times more mass in \hi \ than stars, and its gas reservoir does not appear disturbed at the resolution of our VLA observations (Figure~\ref{fig:HIoverlays}, right). Antlia~B, although associated with a much lower-mass host, is currently quenched, but it has had recent star formation and may resume forming stars in the future. The highly isolated Hedgehog appears red and quenched with no UV detection. This is the opposite of what a simple environmental-quenching picture would predict, and shows that host mass and present-day isolation alone do not determine quenching outcomes. M83-dw14 may have entered the M83 halo too recently for environmental suppression to act, or ram pressure stripping may have been ineffective, perhaps because of a large pericenter passage distance. More generally, orbital history, gas content, and the relevant environmental or internal mechanisms must be considered together to explain why the least isolated system in this comparison remains the most actively star-forming. Unfortunately, it is not feasible to obtain the full 6-dimensional orbital properties of M83-dw14. However, higher resolution and deeper \hi \ observations would provide a clearer indication of whether its gas is beginning to be stripped.

\subsection{Grapes}
In the size-luminosity plane, Grapes' closest analog is Pavo, another isolated, star-forming dwarf galaxy 
\citep{Jones2023,BMP2026}. The two systems occupy broadly similar field environments and show sustained stellar mass growth without prolonged quiescent intervals. Despite these similarities, Grapes retains a larger gas reservoir, whereas Pavo has a comparatively modest \hi\ mass of $\log(M_{\rm HI}/M_\odot)=5.79$ \citep{Jones25}, detectable only through deep targeted observations. 

\subsection{Hydra~B}
Hydra~B is the brightest of our targets, sharing a luminosity similar to that of Kamino \citep{BMP2026}, but with a substantially more extended structure. Like Kamino, it is very isolated, offering a useful comparison between two isolated dwarfs of similar stellar mass. Despite this similarity, the two systems show evidence for different chemical evolution: Hydra~B's RGB spans relatively higher metallicities ([M/H] $=-2.2$ to $-1.2$), while Kamino's RGB is dominated by more metal-poor stars ([M/H]$=-2.5$ to $-2.0$). 
\\

Taken together, these comparisons illustrate that at this low-mass regime, dwarf galaxies exhibit quite diverse evolutionary paths.  This diversity is not unexpected: even among genuinely isolated low-mass galaxies, where environmental effects are minimal or absent, SFHs do not converge on a single characteristic shape \citep{BMP2026}, and are instead driven by a combination of inhomogeneous reionization, stochastic mass-assembly histories, and internal feedback \citep[e.g.,][]{Cole2007,Cole2014,Monelli2010a,Weisz2011,Weisz2014,Skillman2014,Skillman2017,Garrison-Kimmel2019,Digby2019,Joshi2021}.
Given that this diversity persists even among isolated systems alone, it is unsurprising that our sample -- which spans satellites, backsplash candidates, and genuinely isolated dwarfs -- exhibits an equally broad range of SFHs and evolutionary outcomes.

Looking ahead, continued follow-up of other SEAMLESS candidates will be critical to disentangle internal from environmental drivers of quenching at this mass scale. More broadly, upcoming deep, wide-area imaging from facilities such as the Rubin Observatory Legacy Survey of Space and Time \citep{Rubin19}, the Roman Space Telescope \citep{wfirst}, and Euclid \citep{Euclid2025}, combined with machine-learning detection techniques like ours, will substantially expand our ability to detect and characterize faint dwarf galaxies at unprecedented distances \citep[e.g.,][]{RodriguezWimberly2019,BMP2021}. Such efforts will yield robust statistical samples spanning the full range of environments probed here -- isolated, satellite, and backsplash -- enabling more stringent tests of cosmological models on the smallest scales and further clarifying the interplay between environment and internal processes in shaping star formation in low-mass galaxies.

\begin{acknowledgements}
Based on observations made with the NASA/ESA Hubble Space Telescope, obtained at the Space Telescope Science Institute, which is operated by the Association of Universities for Research in Astronomy, Inc., under NASA contract NAS5-26555. All the HST data used in this paper can be found in MAST. These observations are associated with program \# HST-GO-17798, HST-GO-17607, HST-GO-17891, and HST-SNAP-15922. Support for these programs was provided by NASA through a grant from the Space Telescope Science Institute, which is operated by the Association of Universities for Research in Astronomy, Inc., under NASA contract NAS5-26555. 

BMP acknowledges support from NSF grant AST-2508745. DJS and DZ acknowledge support from NSF grants AST-2205863 and 2508746. DC acknowledges support from NSF grant AST-2508747.


\end{acknowledgements}

\vspace{15mm}
\facilities{HST (ACS/WFC3), SOAR, VLA, GBT, Parkes, Blanco}

\software{Astropy \citep{astropy13,astropy18}, The IDL Astronomy User's Library \citep{IDLforever}, \texttt{DOLPHOT} \citep{Dolphin2000}, CASA \citep{CASA}, SoFiA v2 \citep{SoFiA_software}}

\bibliographystyle{aasjournal}
\bibliography{reference}

\clearpage
\appendix
\section{Alternative Names and Catalog Designations}
\label{sec:names}

\begin{table*}[h!] 
\centering
\small
\caption{Adopted names and alternative designations of
the galaxies studied in this work.}
\label{tab:alternative_names}
\begin{tabular}{ll}
\hline
Adopted Name & Alternative Designations \\
\hline
Cetus B [1] & 	SMDG J0040349-203323 [2], Do VIII [3]  \\
Hydrus A [1] & ---  \\
Ursa Major C & SMDG J1333258+563002 [4]   \\
Hydra B & LEDA 166155 [5]  \\
Grapes [7] & SDSS J115205.58+544732.2 [6], SMDG J1152057+544733 [4]   \\
M83-dw14  & ---   \\
Centaurus C (Cen C) & ---   \\
\hline
\end{tabular}
  \begin{tablenotes}
      \small
      \item  We adopt the names in the first column throughout this paper. Alternative designations are taken from previous catalogs and studies. M83-dw14 and Cen~C are reported here for the first time.
      \item[] \textbf{References:}  [1] \citet{Fielder2025}; [2] \citet{Zaritsky2022}; [3] \citet{MD2026}; [4] \citet{Zaritsky2023}; [5] \citet{Hyperleda2003}; [6] \citet{SDSS2009}; [7] \citet{Karachentsev2018}
    \end{tablenotes}
\end{table*}

\FloatBarrier

\section{\hi \ spectrum of Cen~C}
\label{sec:CenC_spec}

\begin{figure}
    \centering
    \includegraphics[width=0.5\linewidth]{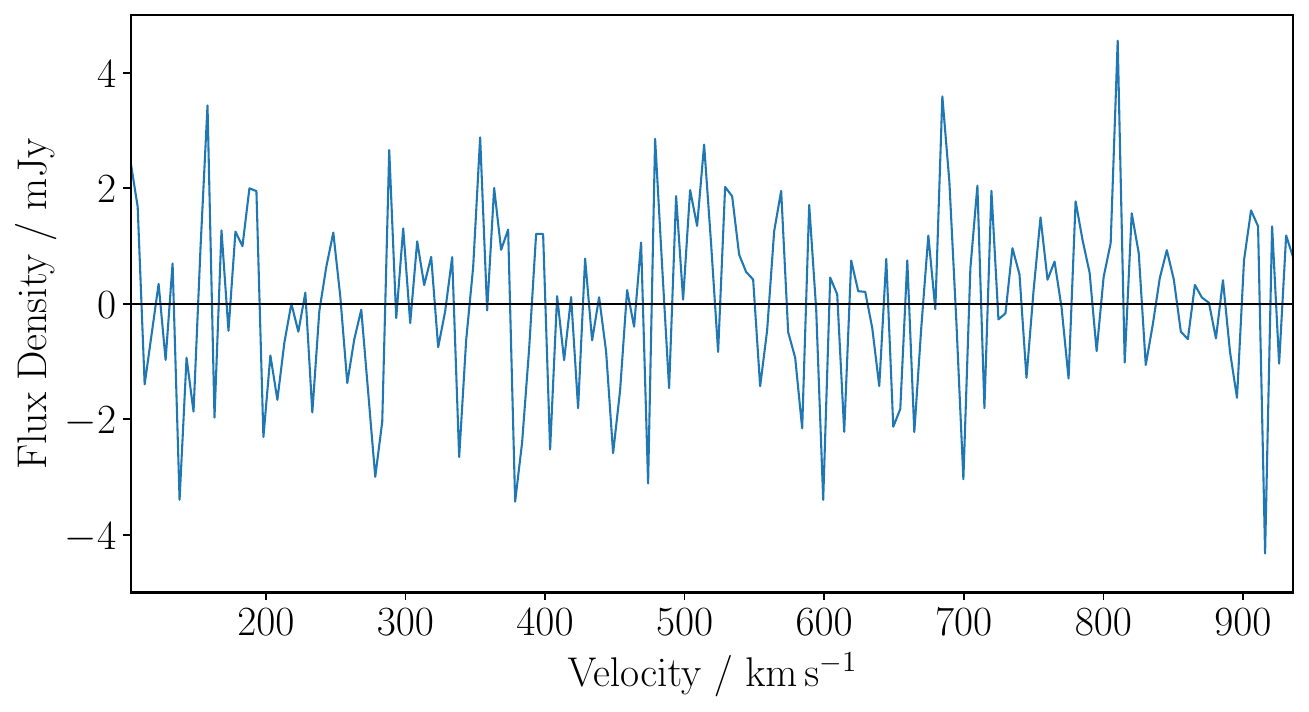}
    \caption{VLA spectrum extracted over one beam area centered on the optical position of Cen~C. The rms noise is 1.54~mJy.}
    \label{fig:CenC_VLA_spec}
\end{figure}

Figure~\ref{fig:CenC_VLA_spec} shows the non-detection of Cen~C in the VLA observation of M83-dw14 (which covered both sources). The spectrum was extracted from the primary beam-corrected data cube, over an area equal to the beam size, centered on the optical position of Cen~C. The rms noise of this spectrum was used to determine the \hi \ mass upper limit in Table~\ref{tab:dwarfs}. Note that the peak flux density of Leo~T is $\sim$0.55~Jy \citep{Adams2018}. If it were at $\sim$4~Mpc, rather than 420~kpc, its peak flux density would be $\sim$6~mJy, which would be readily detectable in the spectrum shown.

As the bandwidth of this observation was limited, spanning only $\sim$800~\kms, the \hi \ limit for Cen~C is only valid if its radial velocity falls within the range shown in Figure~\ref{fig:CenC_VLA_spec}. Without an independent measurement of its radial velocity, there remains the possibility that it in fact has a higher \hi \ mass than our limit in Table~\ref{tab:dwarfs}, if its velocity falls outside the observed range. To ascertain how likely this possibility is we used the CosmicFlows-4 distance--velocity calculator\footnote{\url{http://edd.ifa.hawaii.edu/CF4calculator/}} \citep{Kourkchi2020,Valade2024} to determine the expected radial velocity of Cen~C given its position and distance, obtaining a value of 300~\kms. Thus, to be outside the observed band Cen~C would need a minimum negative peculiar velocity of approximately 200~\kms, or a minimum positive peculiar velocity of over 600~\kms. While neither of these is likely, the former is certainly plausible and cannot be ruled out with our existing observations.

\end{document}